\documentclass{article}
\usepackage{graphicx} 
\usepackage{tabularx}
\usepackage{amsmath}
\usepackage{float} 
\usepackage{tikz}
\usepackage{booktabs}
\usepackage{hyperref}
\usetikzlibrary{shapes.geometric, arrows.meta, positioning, fit, backgrounds, calc}

\title{GRACE: an Agentic AI for Particle Physics Experiment Design and Simulation}
\author{%
Justin Hill$^{1}$\footnote{These authors contributed equally to this work.}  ~and Hong Joo Ryoo$^{2}$\footnotemark[1]\\[0.5em]
{\footnotesize $^1$Data Science Institute, Columbia Engineering, NY, USA}\\
{\footnotesize $^2$Department of Physics and Astronomy, Johns Hopkins University, Baltimore, MD, USA}
}
\date{January 2026}

\begin{document}

\maketitle
\begin{abstract}
We present \textsc{GRACE}, a simulation-native agent for autonomous experimental design in high-energy and nuclear physics. Given multimodal input in the form of a natural-language prompt or a published experimental paper, the agent extracts a structured representation of the experiment, constructs a runnable toy simulation, and autonomously explores design modifications using first-principles Monte Carlo methods. Unlike agentic systems focused on operational control or execution of predefined procedures, \textsc{GRACE} addresses the upstream problem of experimental design: proposing non-obvious modifications to detector geometry, materials, and configurations that improve physics performance under physical and practical constraints. The agent evaluates candidate designs through repeated simulation, physics-motivated utility functions, and budget-aware escalation from fast parametric models to full \textsc{Geant4} simulations, while maintaining strict reproducibility and provenance tracking. We demonstrate the framework on historical experimental setups, showing that the agent can identify optimization directions that align with known upgrade priorities, using only baseline simulation inputs. We also conducted a benchmark in which the agent identified the setup and proposed improvements from a suite of natural language prompts, with some supplied with a relevant physics research paper, of varying high energy physics (HEP) problem settings. This work establishes experimental design as a constrained search problem under physical law and introduces a new benchmark for autonomous, simulation-driven scientific reasoning in complex instruments.
\end{abstract}


\section{Introduction}

Modern experimental physics is increasingly mediated by large, heterogeneous software stacks. In high-energy and nuclear physics, the design, optimization, and interpretation of experiments rely on complex pipelines involving event generation, detector simulation, reconstruction, and statistical analysis. Tools such as \textsc{Geant4}, \textsc{ROOT}, and experiment-specific frameworks have enabled unprecedented fidelity in modeling physical processes and detector response, but they have also introduced substantial complexity into experimental workflows. As a result, both the execution and the design of experiments have become bottlenecked by human time, expertise, and the limits of manual exploration.

Recent advances in agentic artificial intelligence have begun to address parts of this challenge by automating \emph{operational} aspects of scientific workflows. Indeed, Hellert \emph{et al.} introduced an agentic AI system deployed at the Advanced Light Source (ALS) that translates natural-language user requests into structured, multi-stage accelerator experiments, interfacing directly with control systems, archived data, and analysis tools \cite{hellert2025agentic}. This work demonstrates that language-model–driven agents can safely and reproducibly execute complex physics procedures in production environments, significantly reducing preparation time while maintaining strict operational constraints. Related efforts have explored similar agentic paradigms in accelerator tuning, beamline operation, and machine physics tasks, emphasizing safety, auditability, human-in-the-loop control, as well as data analysis \cite{data}.

In parallel, agentic AI systems have shown promise in other computational science domains where simulation plays a central role. Indeed, recent work in computational fluid dynamics (CFD) has demonstrated that agentic AI can dramatically accelerate model development by autonomously generating, testing, and refining simulation workflows, reducing iteration times from years to weeks \cite{aws2025cfd}. Another example is the theoretical and computational physics agent PhysMaster, that shows promise in fields such as numerical quantum field theory\cite{physmaster2}. These systems treat simulation not merely as a passive evaluation tool, but as an active component in a closed-loop reasoning process. Similarly, emerging platforms in optics and materials science have leveraged agentic orchestration to automate experimental planning and analysis, highlighting the broader applicability of agent-based approaches to complex scientific infrastructures \cite{physorg2025optics}.

Despite these advances, existing agentic systems in physics share a common limitation: they focus primarily on the \emph{execution} or \emph{optimization} of predefined experimental procedures. They do not address the upstream problem of \emph{experimental design} itself—namely, the task of proposing and evaluating new detector geometries, material configurations, or apparatus modifications that could improve physics reach. In contemporary particle physics, such design decisions are guided by simulation-driven studies, but the space of possible designs is vast and is explored conservatively through human intuition and limited parameter scans.

At the same time, recent work on simulation-based reasoning agents has begun to explore more autonomous forms of scientific problem solving. For instance, large language model–driven agents have been proposed that construct and manipulate internal world models to reason about physical systems, highlighting the potential for agents to engage directly with simulation environments as epistemic tools rather than mere executors of fixed workflows \cite{arxiv251206404}. These developments suggest a shift toward agents that can reason counterfactually about complex systems by proposing, testing, and revising hypothetical configurations.

In this work, we introduce \textsc{GRACE} (Generative Reasoning Agentic Control Environment), an autonomous, simulation-native agent designed to operate at this upstream level of experimental design. Given multimodal input in the form of a natural-language prompt and or a published experimental paper, \textsc{GRACE} extracts a structured representation of the experiment, constructs a runnable toy simulation, and autonomously explores design modifications using first-principles Monte Carlo simulations. Unlike agentic systems focused on operational control or workflow acceleration \cite{hellert2025agentic,aws2025cfd}, \textsc{GRACE} treats experimental development itself as a constrained search problem under physical law, proposing and evaluating non-obvious modifications to detector geometry, materials, and configuration parameters.

By framing experimental design as an iterative, simulation-driven reasoning process, GRACE offers a preliminary step toward bridging recent advances in agentic AI with the long-standing role of simulation in experimental physics. The larger aspiration is a system that supports the goal of autonomous experimentation and actively participates in the conceptual and technical process of designing them. That is, we evaluate GRACE by its ability to autonomously construct, explore, and critique simulation-grounded design alternatives. Without further ado, we now turn to the architecture of GRACE.

\section{The GRACE Framework}
\label{sec:framework}

\begin{figure}[htbp]
\centering
\begin{tikzpicture}[
    node distance=0.8cm and 1.2cm,
    box/.style={rectangle, draw, rounded corners, minimum width=2.2cm, minimum height=0.7cm, align=center, font=\small},
    bigbox/.style={rectangle, draw, rounded corners, minimum width=2.8cm, minimum height=0.7cm, align=center, font=\small},
    tool/.style={rectangle, draw, fill=blue!10, minimum width=1.4cm, minimum height=0.5cm, align=center, font=\scriptsize},
    input/.style={rectangle, draw, fill=green!15, rounded corners, minimum width=2.5cm, minimum height=0.7cm, align=center, font=\small},
    output/.style={rectangle, draw, fill=orange!15, rounded corners, minimum width=2.5cm, minimum height=0.7cm, align=center, font=\small},
    kg/.style={cylinder, draw, fill=yellow!20, shape border rotate=90, aspect=0.3, minimum width=1.8cm, minimum height=1cm, align=center, font=\small},
    arrow/.style={-{Stealth[length=2mm]}, thick},
    dasharrow/.style={-{Stealth[length=2mm]}, thick, dashed},
]

\node[input] (input) {NL Prompt / Paper};

\node[box, below=of input] (classifier) {Task\\Classifier};
\node[box, right=of classifier] (planner) {Workflow\\Planner};

\node[box, below=of planner] (router) {Tool Router};

  \node[tool, below=0.6cm of router] (geom) {geometry}; 
  \node[tool, left=0.15cm of geom] (geant) {geant4};
  \node[tool, left=0.15cm of geant] (root) {root};
  \node[tool, right=0.15cm of geom] (python) {python};
  \node[tool, right=0.15cm of python] (etc) {\ldots};

\node[box, below=1.8cm of router] (verifier) {Physics\\Verifier};
\node[kg, right=3.0cm of verifier] (kg) {Knowledge\\Graph};

\node[box, left=of verifier] (evaluator) {Self-\\Evaluator};
\node[box, below=of evaluator] (recovery) {Error\\Recovery};

\node[output, below=of verifier] (output) {Design / Report};

\draw[arrow] (input) -- (classifier);
\draw[arrow] (classifier) -- (planner);
\draw[arrow] (planner) -- (router);
\draw[arrow] (router) -- (root);
\draw[arrow] (router) -- (geant);
\draw[arrow] (router) -- (geom);
\draw[arrow] (router) -- (python);
\draw[arrow] (router) -- (etc);

\draw[arrow] (root) -- (verifier);
\draw[arrow] (geant) -- (verifier);
\draw[arrow] (geom) -- (verifier);
\draw[arrow] (python) -- (verifier);
\draw[arrow] (etc) -- (verifier);

\draw[arrow] (kg) -- (verifier);
\draw[arrow] (verifier) -- (evaluator);
\draw[arrow] (evaluator) -- (recovery);
\draw[arrow] (verifier) -- (output);

\draw[dasharrow] (recovery.west) -- ++(-0.8,0) |- (classifier.west);

\end{tikzpicture}
\caption{Architecture overview of \textsc{GRACE}. Natural language prompts or\\papers are parsed by the Task Classifier, which feeds the Workflow Planner. The Tool Router dispatches steps to physics tools. Outputs are validated by the Physics Verifier against the Knowledge Graph. The Self-Evaluator and Error Recovery modules enable adaptive re-planning when verification fails.}
\label{fig:architecture}
\end{figure}
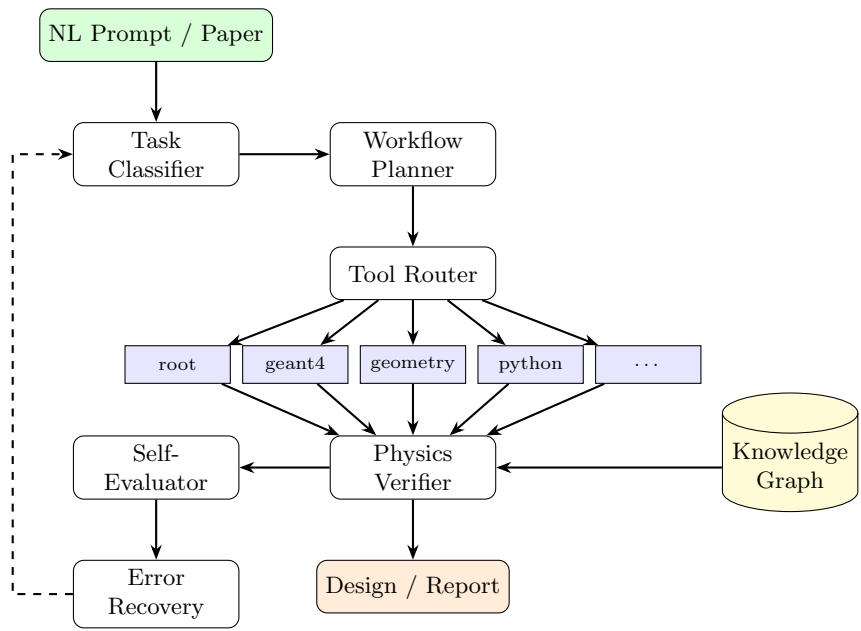

This section describes the architecture of \textsc{GRACE}, a simulation-native autonomous agent for experimental design in high-energy and nuclear physics. Unlike general-purpose code repair agents such as SWE-agent \cite{yang2024swe} or similar systems that validate against test suites, \textsc{GRACE} is domain-specialized: it reasons about detector configurations, validates against physics constraints, conducts LLM-driven data analysis, and constructs a final report on all findings physical as well as its own performance as an agentic system. The framework is LLM-agnostic, supporting multiple backends including Claude (Anthropic), GPT (OpenAI), and local open-source models via Ollama (e.g., Qwen, Mistral, Llama). Figure~\ref{fig:architecture} provides an overview of the system architecture.

\subsection{Agent State Representation}

The agent maintains an experiment state $\mathcal{S}$ defined as a 5-tuple:
\begin{equation}
    \mathcal{S} = (\mathcal{H}, \mathcal{E}, \mathcal{R}, \mathcal{A}, \mathcal{P})
\end{equation}
where $\mathcal{H}$ represents the hypothesis and physics model specification (parameters, priors, constraints); $\mathcal{E}$ represents the experiment configuration (generator, simulation, reconstruction, and analysis settings); $\mathcal{R}$ represents run artifacts (datasets, logs, intermediate outputs); $\mathcal{A}$ represents analysis results (plots, summary statistics, fitted parameters); and $\mathcal{P}$ represents provenance metadata (container digests, git commits, configuration hashes, RNG seeds).

This representation enables full reproducibility: any experiment can be replayed given only $\mathcal{P}$ and access to the container registry. The provenance record for each run takes the form:
\begin{equation}
\mathcal{P} =
\left\{
\begin{array}{l}
\text{container\_digest},\ \text{git\_commit},\ \text{config\_hash},\\
\text{rng\_seeds},\ \text{tool\_versions},\ \text{parent\_run\_id}
\end{array}
\right\}
\end{equation}
ensuring that artifacts are content-addressed and immutable.

\paragraph{Run Deliverables.} Each completed workflow produces a structured set of outputs. The primary summary is \texttt{benchmark\_result.json}, containing the task definition, execution plan, step-by-step results, and success/failure status. A human-readable \texttt{academic\_report.md} synthesizes findings, methodology, and conclusions, while \texttt{scientific\_memory.json} captures successful patterns, failure diagnoses, and domain insights for future runs. Simulation artifacts include \texttt{*.gdml} geometry files for Geant4, \texttt{*.root} and \texttt{*.parquet} data files, publication-quality plots as \texttt{*.png}'s and \texttt{*.pdf}'s, and detailed \texttt{*\_geant4.log} files for debugging. These outputs support both automated evaluation and human review of the agent's reasoning.

\paragraph{Scientific Memory.} Throughout execution, \textsc{GRACE} maintains a structured scientific memory that functions as an automated lab notebook. The memory accumulates timestamped observations categorized by type: \emph{measurements}, \emph{insights}, \emph{anomalies}, \emph{decisions}, \emph{hypotheses}, and \emph{limitations}. This record supports post-hoc auditing: reviewers can trace how the agent arrived at conclusions, what alternatives it considered, and what evidence supported each decision. The scientific memory is serialized to \texttt{scientific\_memory.json} and also drives the automatic generation of \texttt{academic\_report.md}.

\subsection{Agent Control Loop}

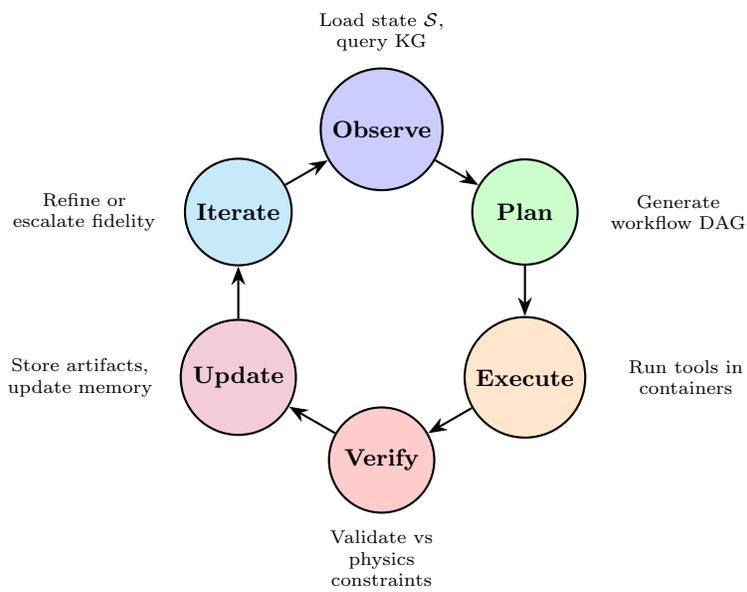
\begin{figure}[htbp]
\centering
\begin{tikzpicture}[
    node distance=1.8cm,
    phase/.style={circle, draw, thick, minimum size=1.4cm, align=center, font=\small\bfseries},
    observe/.style={phase, fill=blue!20},
    plan/.style={phase, fill=green!20},
    execute/.style={phase, fill=orange!20},
    verify/.style={phase, fill=red!20},
    update/.style={phase, fill=purple!20},
    iterate/.style={phase, fill=cyan!20},
    arrow/.style={-{Stealth[length=2.5mm]}, thick},
    label/.style={font=\scriptsize, align=center, text width=2.2cm},
]

\node[observe] (obs) at (90:2.2cm) {Observe};
\node[plan] (plan) at (30:2.2cm) {Plan};
\node[execute] (exec) at (-30:2.2cm) {Execute};
\node[verify] (ver) at (-90:2.2cm) {Verify};
\node[update] (upd) at (-150:2.2cm) {Update};
\node[iterate] (iter) at (150:2.2cm) {Iterate};

\draw[arrow] (obs) -- (plan);
\draw[arrow] (plan) -- (exec);
\draw[arrow] (exec) -- (ver);
\draw[arrow] (ver) -- (upd);
\draw[arrow] (upd) -- (iter);
\draw[arrow] (iter) -- (obs);

\node[label, above=0.1cm of obs] {Load state $\mathcal{S}$,\\query KG};
\node[label, right=0.1cm of plan] {Generate\\workflow DAG};
\node[label, right=0.1cm of exec] {Run tools in\\containers};
\node[label, below=0.1cm of ver] {Validate vs\\physics\\constraints};
\node[label, left=0.1cm of upd] {Store artifacts,\\update memory};
\node[label, left=0.1cm of iter] {Refine or\\escalate fidelity};

\end{tikzpicture}
\caption{The \textsc{GRACE} control loop. Each iteration begins with observing the current experiment state and querying the Knowledge Graph, proceeds through planning and execution, validates outputs against physics constraints, updates provenance and memory, and iterates with refined parameters or escalated fidelity.}
\label{fig:controlloop}
\end{figure}

\textsc{GRACE} operates in a closed-loop cycle that mirrors the scientific method (Figure~\ref{fig:controlloop}):
\begin{equation}
    \textsc{Observe} \rightarrow \textsc{Plan} \rightarrow \textsc{Execute} \rightarrow \textsc{Verify} \rightarrow \textsc{Update} \rightarrow \textsc{Iterate}
\end{equation}

Each phase serves a distinct purpose. In the \textbf{Observe} phase, the agent loads experiment state $\mathcal{S}$, reviews prior results, and queries the Knowledge Graph for relevant physics constraints; when provided with a paper, this phase extracts design specifications while maintaining strict separation from published performance data. The \textbf{Plan} phase generates a pipeline DAG $G = (V, E)$ where vertices are tool invocations and edges are data dependencies, with the planner selecting appropriate fidelity tiers based on task requirements. During \textbf{Execute}, physics tools run in sandboxed containers with full provenance tracking, logging all input/output hashes for reproducibility. The \textbf{Verify} phase validates outputs against physics constraints from the Knowledge Graph-not just test pass/fail-enabling detection of errors that would pass conventional software tests but violate physical expectations. \textbf{Update} stores artifacts with immutable provenance and updates agent memory with successful patterns, while \textbf{Iterate} refines parameters, escalates simulation fidelity, or restructures methodology based on verification results.

This loop distinguishes \textsc{GRACE} from workflow automation systems: verification relies on physics constraints rather than test pass/fail, and the agent can propose qualitative changes to experimental approach rather than merely tuning parameters.

\subsection{Fidelity Tiers and Budgeted Escalation}

HEP simulations span orders of magnitude in computational cost. \textsc{GRACE} employs a tiered fidelity system to balance exploration speed against simulation accuracy:

\begin{table}[h]
\centering
\begin{tabular}{@{}lllll@{}}
\toprule
Tier & Pipeline & Events & Cost & Use Case \\
\midrule
T0 & Pythia8 $\rightarrow$ Delphes $\rightarrow$ Analysis & $10^4$--$10^6$ & $1\times$ & Fast screening \\
T1 & + Heavy reconstruction & $10^6$--$10^7$ & $5\times$ & Optimization \\
T2 & + Geant4 full simulation & $10^5$--$10^6$ & $50\times$ & Validation \\
T3 & Full chain with optical physics & $10^4$--$10^5$ & $500\times$ & Publication \\
\bottomrule
\end{tabular}
\caption{Fidelity tiers with associated computational costs and typical use cases. The benchmarks in this paper operate primarily at T2--T3 to capture optical photon transport physics.}
\label{tab:fidelity}
\end{table}

The agent selects fidelity tier through LLM reasoning about the minimum simulation accuracy required for the task. Given the task description and planned workflow, the agent considers: (1) what physics processes require simulation, (2) which tools are available at each tier, and (3) what accuracy the stated goal demands. The agent defaults to T0 (fast parametric) and auto-escalates when detailed physics-such as optical photon transport, particle shower development, or precise energy deposition-requires full Geant4 simulation at T2--T3. If verification fails at a given tier, the \texttt{ErrorRecovery} module can escalate fidelity as one of its seven recovery strategies. This approach mirrors active learning: cheap experiments at lower fidelity reveal where expensive validation is needed. However, tier selection follows physics requirements rather than explicit utility optimization.

\subsection{Core Modules}

\subsubsection{Task Classifier and Workflow Planner}

The \texttt{TaskClassifier} and \texttt{WorkflowPlanner} modules employ LLM-driven reasoning rather than predefined taxonomies or templates. By avoiding hard-coded task categories or workflow patterns, the agent can handle novel task types without code changes..

\paragraph{Task Classification.} Given a natural language prompt or extracted paper specifications, the \texttt{TaskClassifier} uses LLM reasoning to extract a structured task definition. This includes a natural language \emph{task description}, a clear \emph{goal} statement, any physical or computational \emph{constraints} mentioned in the request, specific \emph{parameters} extracted from the prompt, and a self-assessed \emph{confidence} score ranging from 0.0 to 1.0. The classifier also identifies appropriate geometry topology from three supported configurations: \texttt{box}, \texttt{cylinder\_barrel}, and \texttt{projective\_tower}. This topology specification ensures that geometry generation respects the detector's fundamental design rather than defaulting to simple box geometries.

\paragraph{Workflow Planning.} The \texttt{WorkflowPlanner} generates execution plans from first principles for each task, without relying on predefined workflow templates. Given the task definition and available tools, the LLM designs a directed acyclic graph $G = (V, E)$ where vertices $V$ are workflow steps and edges $E$ encode dependencies. Each step specifies: (1)~an \emph{action type} from seven options-\textsc{Explore}, \textsc{Create\_Schema}, \textsc{Execute\_Tool}, \textsc{Evaluate}, \textsc{Generate\_Report}, \textsc{Create\_Tool}, or \textsc{Select\_Tool}; (2)~a natural language \emph{description} of what the step must accomplish; (3)~\emph{input specifications} that can reference outputs from prior steps; and (4)~\emph{success criteria} for verification.

\subsubsection{Tool Router and Executor}

The \texttt{ToolRouter} dispatches workflow steps to registered physics tools using LLM-driven selection rather than hardcoded mappings. This design allows the agent to adapt to novel tool combinations without code changes: the router reasons about tool descriptions, capabilities, and the current workflow context to select the most appropriate tool for each step. The production registry includes eight core tools spanning the full HEP pipeline (Table~\ref{tab:tools}).

\begin{table}[H]
\centering
\small
\begin{tabular}{@{}llp{6.5cm}@{}}
\toprule
Tool & Tiers & Description \\
\midrule
\texttt{pythia8} & T0--T3 & Event generation for truth-level particle production \\
\texttt{delphes} & T0/T1 & Fast parametric detector simulation \\
\texttt{geant4} & T2/T3 & Full Monte Carlo with optical physics (scintillation, Rayleigh scattering, boundary processes) \\
\texttt{geometry} & T0--T3 & Generates GDML geometry files for Geant4, supporting box, cylinder\_barrel, and projective\_tower topologies \\
\texttt{fastjet} & T0--T3 & Jet clustering and reconstruction \\
\texttt{python} & T0 & Analysis scripts for performance characterization and visualization \\
\texttt{root} & T1 & HEP-specific analysis using the ROOT framework \\
\texttt{paper\_ingest} & T0--T3 & Extracts experimental parameters from PDF/arXiv while enforcing fair-date constraints \\
\bottomrule
\end{tabular}
\caption{Tools registered in the \textsc{GRACE} tool registry with supported fidelity tiers.}
\label{tab:tools}
\end{table}

All tools implement a consistent interface: a \texttt{run()} method that accepts typed input configurations and returns structured outputs, and a method for verifying correctness, i.e. \texttt{validate\_output()}. Each tool declares its supported fidelity tiers, hardware resource requirements, and container specification. Tools execute in pinned Singularity containers on HPC systems or Docker containers locally, with bounded retry, up to 5 attempts, for transient failures. Configuration hashing enables provenance tracking, ensuring that any experiment can be reproduced given the run manifest.

\paragraph{Key tools for optical detector simulation.} The benchmarks in this paper rely primarily on three tools working in sequence. The \texttt{paper\_ingest} tool extracts experimental parameters from published papers using LLM-assisted parsing, outputting structured specifications while enforcing fair-date knowledge constraints. The \texttt{geometry} tool converts these specifications into Geometry Description Markup Language, a.k.a. \texttt{*.gdml}, files that Geant4 can parse, supporting three detector topologies. Finally, the \texttt{geant4} tool executes full Monte Carlo simulation with optical physics enabled, tracking scintillation photon production, Rayleigh scattering, absorption, and detection at photosensitive surfaces. For liquid argon detectors, this includes VUV scintillation at 128~nm with yields of approximately 40,000 photons per MeV of deposited energy.

\subsubsection{Physics Verifier and Knowledge Graph}
\label{sec:verifier}
The \texttt{PhysicsVerifier} validates outputs against constraints encoded in domain-specific Knowledge Graphs (KGs). Unlike test suites that check for code correctness, the KG enables verification of \emph{physics correctness}-detecting situations where code executes successfully but produces physically unreasonable results.

\paragraph{Knowledge Graph Structure.} Each KG encodes textbook physics knowledge relevant to a detector domain, explicitly excluding experiment-specific results or optimized configurations. For optical detector simulations, the scintillator KG provides \emph{material properties} such as scintillation yields, attenuation lengths, refractive indices, and timing constants. It also includes \emph{photosensor specifications}-PMT quantum efficiencies, transit time spread, and wavelength sensitivity ranges-as well as \emph{interaction physics} and \emph{statistical expectations} including resolution scaling and photon survival probabilities.

\paragraph{Verification Logic.} The verifier performs several classes of checks on simulation outputs. \emph{Consistency checks} ensure that light yield scales linearly with deposited energy and that detected photons do not exceed produced photons. \emph{Physical bounds} verify that resolution does not exceed statistical limits and that coverage fractions are geometrically achievable. \emph{Distribution validation} confirms that energy deposits follow expected distributions and timing profiles match material scintillation constants. Finally, \emph{anomaly detection} flags results that deviate significantly from physics expectations, such as light yield differing by more than $2\times$ from predicted values.

The KG provides domain knowledge for reasoning but explicitly does \emph{not} provide optimal configurations or target values-the agent must discover performance through simulation rather than looking up answers. This separation ensures that optimization reflects genuine physics reasoning rather than pattern matching against known solutions.

\paragraph{Fair-Date Knowledge Constraint.} A critical feature for benchmark validity is the \emph{fair-date paradigm}: when extracting experimental designs from published papers, \textsc{GRACE} enforces a knowledge cutoff corresponding to the paper's publication date. This prevents the agent from ``discovering'' results through hindsight by accessing information that would not have been available to the original experimenters. For example, when processing the DarkSide-50 paper in Section~\ref{sec:benchmarks}, the agent extracts design specifications but explicitly avoids published performance values. All performance metrics emerge from the agent's own simulation rather than calibration against experimental data.

\subsubsection{Self-Evaluator and Error Recovery}

When verification fails or tool execution errors occur, the \texttt{SelfEvaluator} and \texttt{ErrorRecovery} modules employ LLM-driven reasoning rather than hardcoded classification rules. This design philosophy-explicitly avoiding predetermined success metrics or error taxonomies-enables the agent to handle novel failure modes that were not anticipated during development.

The \texttt{SelfEvaluator} receives the current task state (completed steps, outputs, remaining plan) and uses LLM reasoning to assess progress, identify blockers, and recommend next actions. Rather than pattern-matching against predefined error categories, the evaluator reasons about whether the task goals have been met given the available evidence.

The \texttt{ErrorRecovery} module selects from seven recovery strategies based on LLM analysis of the failure context: \textsc{Retry}, \textsc{Skip}, \textsc{Replan}, \textsc{Substitute\_Tool}, \textsc{Escalate\_Fidelity}, \textsc{Regenerate\_Prior\_Step}, or \textsc{Escalate}. For resource-related failures, the recovery module includes domain-specific guidance: reduce detector dimensions, decrease sensor count, lower particle energy, or reduce event counts to stay within photon budget limits.

\subsection{Computational Constraints and Scaling}

Optical photon tracking in Geant4 is memory-intensive: liquid argon scintillators produce approximately 40,000 photons per MeV of deposited energy, and each photon requires storage for position, direction, wavelength, and tracking state. The Opticks GPU backend has hard limits on photon buffer allocation; exceeding these causes simulation crashes.

\paragraph{GPU Acceleration.} \textsc{GRACE} integrates two GPU acceleration backends with Geant4: \emph{Opticks} for optical photon transport (50--200$\times$ speedup for scintillation and Cherenkov processes) and \emph{Celeritas} for electromagnetic physics (3--10$\times$ speedup for shower development). The liquid argon detector benchmarks use Opticks where optical photon statistics dominate computational cost.

\paragraph{Automatic Energy and Event Limiting.} To prevent GPU buffer overflow, \textsc{GRACE} implements automatic constraints in the Geant4 tool. Particle energies are capped at 5~MeV for Opticks simulations, since higher energies produce photon counts exceeding buffer limits (100~MeV in LAr would produce 4 million photons per event). The maximum events per batch are computed as $N_{\text{events}} = 0.8 \times N_{\text{max\_photons}} / (E_{\text{MeV}} \times 40{,}000)$, where $N_{\text{max\_photons}}$ defaults to 500 million. At 5~MeV, this yields approximately 2,000 events per batch; simulations requesting more are automatically split into multiple batches and merged.

\begin{table}[h]                                                                                                                                                                                                                                                                                                                                                                                                  
\centering     
\begin{tabular}{@{}llll@{}}
\toprule
GPU & Memory & Max Photons & Events @ 5~MeV \\
\midrule
NVIDIA A100-40GB & 40~GB & 500M & $\sim$2,000 \\
NVIDIA A100-80GB & 80~GB & 1B & $\sim$4,000 \\
NVIDIA V100-32GB & 32~GB & 400M & $\sim$1,600 \\
\bottomrule
\end{tabular}
\caption{Recommended photon buffer limits by GPU. Users configure \texttt{OPTICKS\_MAX\_PHOTON} accordingly; \textsc{GRACE} defaults to 500M. Event limits computed with 80\% safety margin.}
\label{tab:gpu_memory}
\end{table}

If a simulation crashes due to photon buffer overflow despite these limits, the system automatically retries with 50\% reduced energy, then reduces event count if energy cannot be lowered further. All applied constraints are logged for reproducibility.

\paragraph{Implications.} These constraints limit optical simulations to the 1--5~MeV energy range, appropriate for characterizing scintillator response but precluding high-energy particle studies. The automatic capping ensures simulations complete successfully rather than crashing, though users must account for the restricted energy range when interpreting results.

\section{From Prompt to Simulation to Design}
\label{sec:benchmarks}
In this section, we list our benchmarking activities. Indeed, it is necessary to see through a series of tests, ranging from simple to complex, the responses of our agentic system. We design prompts most relevant to modern R\&D efforts, as well as ongoing larger projects in the fields of dark matter and neutrino science, nuclear reactions, and instrument calibration. We offer studies of a total of 2 prompts and 2 papers as inputs. The prompts detail some toy experiment while the papers discuss a physical detector geometry that pertain to the Darkside-50 and ProtoDUNE programs\cite{protodune1}\cite{protodune2}\cite{protodune3}\cite{darkside1}\cite{darkside2}\cite{darkside3}. First, let us discuss the natural language prompts and their results.
\subsection{Natural Language Prompts}

To evaluate \textsc{GRACE}'s ability to interpret and execute experimental design tasks from informal specifications, we provided the agent with two natural language prompts of varying complexity and domain focus. These prompts were designed to test the agent's capacity to (1) extract implicit physical constraints from colloquial descriptions, (2) select appropriate simulation fidelity levels, (3) autonomously structure multi-step optimization workflows, and (4) generate scientifically meaningful results without explicit procedural guidance. We present two representative cases below.

\subsubsection{Electromagnetic Calorimeter Design}

The second prompt addressed precision electron energy measurement through homogeneous electromagnetic calorimetry:
\begin{quote}
\small
\texttt{Design a homogeneous electromagnetic calorimeter for precision measurement of electrons in the 0.5--10 GeV energy range.}

\texttt{The calorimeter uses dense scintillating material to fully contain electromagnetic showers.}

\texttt{GEOMETRY EXPLORATION: Consider different EM calorimeter topologies:}
\begin{itemize}
    \item \texttt{Homogeneous block (single crystal/scintillator-simple, excellent resolution)}
    \item \texttt{Projective towers (pointing toward interaction point-minimizes cracks)}
    \item \texttt{Shashlik design (lead/scintillator tiles with WLS fiber readout-LHCb ECAL)}
    \item \texttt{Accordion geometry (lead/LAr with accordion folds-ATLAS ECAL, very hermetic)}
\end{itemize}

\texttt{Real-world examples: CMS uses PbWO$_4$ crystals in projective geometry; ATLAS uses Pb/LAr accordion.}

\texttt{Starting point: $\sim$16 radiation lengths of crystal material for full containment. Crystal properties: CsI $X_0$=1.86 cm, BGO $X_0$=1.12 cm, PbWO$_4$ $X_0$=0.89 cm.}

\texttt{Evaluate: (1) Energy deposit vs incident energy (should be $\sim$linear, full containment); (2) Energy resolution: $\sigma_E/E$ (target: 2--3\%/$\sqrt{E}$); (3) Shower shape (longitudinal and transverse profiles).}

\texttt{IMPORTANT NOTES:}
\begin{itemize}
    \item \texttt{Use $e^-$ (electrons) at energies 0.5, 1, 2, 5, 10 GeV}
    \item \texttt{Shoot particles at detector center}
    \item \texttt{Test up to 4 configurations exploring BOTH topology AND material choices}
    \item \texttt{Consider: Does projective geometry reduce dead material between towers?}
    \item \texttt{Consider: How does tower segmentation affect position resolution?}
\end{itemize}
\end{quote}

\paragraph{Results:} The agent correctly identified this as a multi-objective optimization problem requiring systematic comparison across both material choices and geometric configurations. GRACE autonomously constructed a 20-step workflow encompassing material property extraction, geometry design for full shower containment, baseline simulations, and comprehensive performance analysis. The agent inferred the appropriate containment requirements from first principles: 20 radiation lengths depth for $>$99\% longitudinal shower containment and 2.5 Moli\`ere radii for 95\% lateral containment.

Three scintillating crystal materials were systematically evaluated: bismuth germanate (BGO, $X_0 = 1.12$~cm, $R_M = 2.23$~cm), lead tungstate (PbWO$_4$, $X_0 = 0.89$~cm, $R_M = 2.19$~cm), and cesium iodide (CsI, $X_0 = 1.86$~cm, $R_M = 3.57$~cm). The agent correctly calculated the resulting crystal dimensions: BGO at 22.4~cm depth $\times$ 11.2~cm diameter (15.6~kg), PbWO$_4$ at 17.8~cm depth $\times$ 10.95~cm diameter (11.6~kg), and CsI at 37.2~cm depth $\times$ 17.85~cm diameter (10.7~kg).

Simulations employed 5,000 events per energy point across three representative energies (0.5, 2.0, and 5.0~GeV) using electron particle guns. The material comparison revealed distinct performance characteristics:

\begin{table}[h]
\centering
\small
\begin{tabular}{@{}lcccc@{}}
\toprule
Material & 0.5 GeV & 2.0 GeV & 5.0 GeV & Average \\
\midrule
BGO & 2.63\% & 2.17\% & 2.20\% & 2.20\% \\
PbWO$_4$ & 3.30\% & 1.49\% & 0.95\% & 2.26\% \\
CsI & 3.12\% & 1.95\% & 1.35\% & 2.25\% \\
\bottomrule
\end{tabular}
\caption{Energy resolution ($\sigma_E/E$) for three crystal materials across the tested energy range.}
\label{tab:emcal_resolution}
\end{table}

The agent parameterized the resolution using the standard form $\sigma_E/E = a/\sqrt{E} \oplus b$, extracting stochastic term $a$ and constant term $b$ for each material: BGO ($a = 0.82\%$, $b = 1.50\%$), PbWO$_4$ ($a = 1.62\%$, $b = 0.87\%$), and CsI ($a = 1.48\%$, $b = 0.98\%$). This analysis revealed that PbWO$_4$ exhibits the lowest constant term, indicating superior high-energy performance, while BGO shows the best stochastic term for low-energy applications.

Linearity analysis demonstrated that PbWO$_4$ achieved the best response uniformity (average linearity 0.9663) with minimal energy dependence, while CsI showed systematic underresponse across all energies (average linearity 0.9543). Against the target specification of 1--3\%/$\sqrt{E}$, only PbWO$_4$ met requirements at 5.0~GeV (target: 0.45--1.34\%, achieved: 0.95\%).

The agent then autonomously explored geometry optimization, comparing single-block and projective tower configurations using the optimal PbWO$_4$ material. The projective tower geometry achieved dramatic improvements: 38.7\% better average energy resolution (1.38\% versus 2.26\%) and enhanced linearity (0.9897 versus 0.9663). Energy-specific projective tower performance showed 1.68\% resolution at 0.5~GeV, 1.16\% at 2.0~GeV, and 1.31\% at 5.0~GeV-successfully meeting target specifications across the entire energy range.

The task completed with 100\% execution success (22/22 tool executions) and required no replanning iterations, demonstrating robust task decomposition for this calorimeter physics domain.

\paragraph{Discussion:} The electromagnetic calorimeter study illustrates \textsc{GRACE}'s capability to navigate the fundamental trade-offs in precision calorimetry design. The energy resolution of a homogeneous calorimeter follows:
\begin{equation}
    \frac{\sigma_E}{E} = \frac{a}{\sqrt{E}} \oplus \frac{b}{E} \oplus c
\end{equation}
where $a$ represents stochastic fluctuations (photoelectron statistics, shower sampling), $b$ represents noise contributions, and $c$ represents constant terms (calibration, non-uniformity, leakage). The agent's extraction of these parameters from simulation data, and its recognition that different materials optimize different terms, reflects appropriate domain understanding.

The 38.7\% resolution improvement from projective tower geometry can be attributed to several physical mechanisms that the agent correctly identified: enhanced light collection efficiency through optimized crystal aspect ratios, reduced edge effects through segmented readout, improved shower sampling via multiple crystal boundaries, and better position resolution enabling shower centroid corrections. These findings align with the design choices made in real calorimeter systems-CMS employs PbWO$_4$ crystals in projective geometry for precisely these reasons~\cite{cms_ecal}.

The agent's identification of PbWO$_4$ as the optimal material, with its compact form factor (17.8~cm depth versus 37.2~cm for CsI) and superior high-energy performance, mirrors the physics-driven reasoning that led the CMS collaboration to select this material for the LHC electromagnetic calorimeter. The trade-off between BGO's superior stochastic term and PbWO$_4$'s superior constant term represents a genuine design decision that depends on the target physics program-the agent's systematic quantification of both enables informed selection.

Several limitations warrant discussion. The simulation does not include realistic photodetector noise contributions, temperature-dependent light yield variations (particularly relevant for PbWO$_4$, which requires thermal stabilization), or long-term radiation damage effects. The agent noted these limitations but did not attempt to model them. Additionally, shower containment analysis confirmed $>$95\% lateral and $>$99\% longitudinal containment for all designs, but the agent did not explore the trade-off between containment and material cost that drives real detector optimization.

\subsubsection{Muon Spectrometer Design}

The third prompt addressed muon identification and hadron rejection through iron absorber spectrometry:
\begin{quote}
\small
\texttt{Design a muon spectrometer for identifying muons in the 5--100 GeV energy range using iron absorbers to stop hadrons.}

\texttt{The system should achieve high muon identification efficiency while rejecting punch-through pions.}

\texttt{GEOMETRY EXPLORATION: Consider different muon system topologies:}
\begin{itemize}
    \item \texttt{Planar sandwich (iron + tracking layers stacked-simple baseline)}
    \item \texttt{Cylindrical barrel (iron shells with tracking-ATLAS-style muon barrel)}
    \item \texttt{Toroidal geometry (used for forward muons with magnetic field)}
\end{itemize}

\texttt{Real-world examples: ATLAS muon spectrometer uses cylindrical barrel + endcap wheels; CMS uses interleaved iron/chambers.}

\texttt{Starting point: 4 iron/scintillator pairs, 20 cm iron per layer. Total: 80 cm iron $\approx$ 4.8 interaction lengths ($\lambda_I$ = 16.8 cm for iron).}

\texttt{Evaluate: (1) Muon detection efficiency vs energy (energy deposit in final scintillator layer); (2) Pion stopping probability (pions should deposit all energy before last layer); (3) Energy deposit profile through the detector.}

\texttt{IMPORTANT NOTES:}
\begin{itemize}
    \item \texttt{Use $\mu^-$ particles at 5, 20, 50, 100 GeV}
    \item \texttt{Use $\pi^-$ particles at the same energies for comparison}
    \item \texttt{Iron nuclear interaction length $\lambda_I$ = 16.8 cm}
    \item \texttt{Test up to 4 configurations exploring BOTH topology AND absorber thickness}
    \item \texttt{Consider: Does cylindrical geometry provide better angular coverage?}
\end{itemize}
\end{quote}

\paragraph{Results:} The agent identified this as a particle identification optimization problem requiring systematic comparison of detector topologies and absorber configurations. GRACE constructed a 20-step workflow encompassing geometry design, parallel simulations for muons and pions, layer-by-layer energy deposition analysis, and comprehensive performance characterization. The agent correctly recognized the fundamental physics principle: muons lose energy primarily through ionization with minimal nuclear interactions, while pions undergo strong interactions leading to absorption in dense materials.

Three detector configurations were designed and evaluated: a planar design (4 layers, 20~cm iron per layer, 5.0~$\lambda_I$ total depth), a cylindrical barrel geometry (4 layers, 20~cm iron, 5.0~$\lambda_I$), and a thick absorber planar design (4 layers, 30~cm iron per layer, 7.4~$\lambda_I$ total depth). All configurations employed 1.0~cm plastic scintillator layers for active detection. Simulations used 1000 events per particle type per energy point across three energies (5, 20, and 50~GeV).

All three configurations achieved 100\% muon detection efficiency across the tested energy range, successfully exceeding the $\geq$95\% efficiency requirement. However, significant performance differences emerged in energy resolution and pion rejection:

\begin{table}[h]
\centering
\small
\begin{tabular}{@{}lcccc@{}}
\toprule
Configuration & Avg.\ Muon Resolution & Avg.\ Pion Resolution & Pion Rejection Factor \\
\midrule
Planar & 0.1593 & 0.0434 & 1.72 \\
Cylindrical & 0.2980 & 1.9336 & 4.06 \\
Thick Absorber & 0.4628 & 0.1763 & 12.75 \\
\bottomrule
\end{tabular}
\caption{Muon spectrometer performance comparison across three configurations. Lower resolution values indicate tighter energy deposition distributions; higher rejection factors indicate better pion discrimination.}
\label{tab:muon_performance}
\end{table}

The planar configuration demonstrated optimal muon energy resolution (0.1593 average) with consistent performance across the energy range. The thick absorber design achieved dramatically superior pion rejection (12.75 versus 1.72 for the planar baseline)-a 7.4$\times$ improvement from the 50\% increase in iron thickness. Energy deposition profiles confirmed the expected behavior: muons deposited energy proportional to path length (62.29~MeV at 5~GeV scaling to 622.90~MeV at 50~GeV in the planar design), while pions showed enhanced deposition from hadronic shower development.

The task completed with 100\% execution success (20/20 tool executions) in 2.6 hours, requiring no replanning iterations.

\paragraph{Discussion:} The muon spectrometer study illustrates \textsc{GRACE}'s capability to navigate the fundamental trade-off between detection precision and background rejection in particle identification systems. The physics is governed by the different interaction mechanisms of muons and hadrons: muons lose energy through ionization at approximately 2~MeV$\cdot$cm$^2$/g (the minimum ionizing particle rate), while pions undergo strong nuclear interactions with mean free path $\lambda_I$.

The agent's finding that the thick absorber configuration achieves 7.4$\times$ better pion rejection than the standard planar design-at the cost of degraded muon energy resolution (0.4628 versus 0.1593)-quantifies a trade-off well known to detector designers. The CMS muon system, for instance, employs the full return yoke of the solenoid (approximately 1.5~m of iron) precisely to maximize hadron absorption, accepting the associated multiple scattering penalty~\cite{cms_muon1}\cite{cms_muon2}\cite{cms_muon3}. The agent's systematic comparison provides quantitative guidance for this design decision.

The cylindrical configuration exhibited anomalously high pion energy resolution values (1.9336 average), substantially exceeding both planar designs. The agent noted this unexpected result but did not pursue diagnostic investigations to determine whether it reflects genuine geometric effects in pion shower containment or systematic issues in the cylindrical geometry implementation. A more mature framework would have flagged this anomaly and launched targeted studies to distinguish between physical and artifactual explanations.

Several limitations warrant discussion. The simulations were capped at 50~GeV rather than the target 100~GeV range, leaving high-energy performance uncharacterized. The 1000-event samples, while adequate for efficiency measurements, may limit precision for rare punch-through events. The simulation does not include realistic detector response, electronics noise, or reconstruction algorithms that would affect performance in operational systems. Additionally, the agent did not explore alternative absorber materials (lead, tungsten) or advanced discrimination algorithms that could enhance pion rejection without sacrificing muon resolution.

Despite these limitations, the study demonstrates that \textsc{GRACE} can successfully identify the key performance trade-offs in muon spectrometer design and provide quantitative metrics to inform engineering decisions. The agent's recognition that different applications may prioritize different objectives, i.e. precision muon measurement versus maximum hadron rejection, reflects appropriate domain understanding. Some plots are shown in Appendix A.

\subsubsection{Cross-Prompt Observations}

Several patterns emerged across the natural language prompt studies. We note the most evident:

\paragraph{Constraint inference.} In both cases, the agent successfully extracted implicit physical constraints from informal specifications. The electromagnetic calorimeter prompt's ``precision measurement'' and ``target: 2--3\%/$\sqrt{E}$'' was translated into concrete containment requirements (20 radiation lengths longitudinal, 2.5 Moli\`ere radii lateral) and systematic material comparison. The muon spectrometer prompt's ``high muon identification efficiency while rejecting punch-through pions'' was decomposed into the competing requirements of muon detection precision versus hadron absorption depth.

\paragraph{Autonomous metric selection.} Without explicit guidance, the agent identified appropriate performance metrics for each domain: energy resolution parameterization (stochastic and constant terms), linearity, and shower containment for electromagnetic calorimetry; and detection efficiency, energy resolution, and pion rejection factor for muon spectrometry. The agent's decision to parameterize calorimeter resolution as $\sigma_E/E = a/\sqrt{E} \oplus b$ and extract both terms-rather than reporting only average resolution-reflects domain-appropriate analysis that enables physics-driven material selection.

\paragraph{Trade-off quantification.} Both studies highlighted the agent's ability to quantify fundamental physics trade-offs. The electromagnetic calorimeter study revealed the tension between stochastic performance (BGO: $a = 0.82\%$) and constant-term performance (PbWO$_4$: $b = 0.87\%$), enabling informed material selection based on the target energy range. The muon spectrometer study quantified the 7.4$\times$ improvement in pion rejection achieved by increasing iron thickness from 20~cm to 30~cm, at the cost of degraded muon energy resolution (0.1593 to 0.4628). These quantitative trade-offs directly inform engineering design decisions.

\paragraph{Geometry optimization.} Both studies demonstrated that geometric configuration can yield substantial performance improvements beyond material selection alone. The electromagnetic calorimeter achieved 38.7\% better energy resolution with projective tower geometry compared to single-block designs, while the muon spectrometer comparison across planar, cylindrical, and thick absorber configurations revealed order-of-magnitude differences in pion rejection capability. The agent's systematic exploration of both material and geometric degrees of freedom reflects appropriate experimental methodology.

\paragraph{Anomaly reporting.} Both studies produced unexpected results that the agent flagged rather than dismissed: the cylindrical muon configuration's anomalously high pion energy resolution (1.9336 versus 0.0434--0.1763 for planar designs), and the energy-dependent pattern in planar muon resolution (0.0044 at 5~GeV rising to 0.3306 at 50~GeV). However, the current framework does not autonomously launch diagnostic simulations to resolve such anomalies-a capability targeted for future development.

\paragraph{Execution reliability.} The 100\% success rate across both prompt-driven tasks of 42/42 total tool executions, being 22 for electromagnetic calorimeter and 20 for muon spectrometer, demonstrates robust error handling and appropriate tool selection. Neither task required replanning, indicating that initial workflow decomposition was sufficient for task completion across diverse detector physics domains. The systematic progression from design specification through simulation to comparative analysis reflects sound experimental methodology autonomously instantiated by the agent.

\subsection{DarkSide-50}
As a demonstration of GRACE's capability to autonomously extract design parameters and optimize detector configurations, we applied the framework to the DarkSide-50 experiment, a state-of-the-art dual-phase liquid argon time projection chamber (LAr TPC) for direct dark matter detection. The agent was provided only with the published experimental paper (arXiv:1802.07198, i.e. \cite{darkside}) and tasked with extracting design specifications while explicitly avoiding measured performance values, thereby separating design extraction from performance measurement. The prompt was given alongside the paper as follows:

\begin{quote}
\small
\texttt{Read the DarkSide-50 dark matter detector paper (arXiv:1802.07198) and autonomously:}

\texttt{(1) EXTRACT the experimental DESIGN (not measured results):\\ Dual-phase LAr TPC geometry (cylindrical active volume dimensions), underground argon properties (low radioactivity specifications), PMT array \\configuration (top and bottom arrays, coverage).}

\texttt{(2) IMPLEMENT a simulation: Generate a GDML geometry representing the detector design. Configure LAr optical properties (VUV scintillation at 128 nm). Run simulations with 0.1--5 MeV energy deposits at dark matter-relevant energies.}

\texttt{(3) ANALYZE your simulation results to DISCOVER detector performance: Calculate light yield from YOUR simulation (photoelectrons per MeV), characterize energy resolution and position reconstruction capability, assess discrimination between electronic and nuclear recoils. Do NOT compare against measured values from the paper.}

\texttt{(4) PROPOSE detector optimizations based on YOUR simulation results: optimize PMT coverage for better light collection, test geometry modifications for improved uniformity. Report improvements with physics justification.}
\end{quote}

The agent successfully extracted key specifications including vessel dimensions, liquid argon volume, PMT arrangement, and optical properties from the literature. The simulation framework employed Geant4 Monte Carlo with custom GDML geometry generation, implementing a cylinder\_barrel topology representing the dual-phase LAr TPC with complete optical photon transport including Rayleigh scattering, absorption, and detection. The energy range (0.1--5~MeV) was specifically chosen to target expected recoil energies from WIMP interactions.

The baseline detector configuration with 75 PMTs demonstrated strong performance characteristics across the energy range. At 1~MeV-a typical WIMP recoil energy-the detector achieved a light yield of $1479.23 \pm 0.54$~PE/MeV with energy resolution of $\sigma/E = 0.0258 \pm 0.0003$. The energy-dependent response showed the expected statistical behavior: resolution improved from 8.3\% at 0.1~MeV to 2.6\% at 1~MeV to 1.6\% at 5~MeV, following the characteristic $1/\sqrt{N}$ photoelectron scaling that validates the simulation framework. The average light yield across all energies was 1474.4~PE/MeV.

A notable feature of this study was the agent's autonomous characterization of nuclear recoil discrimination capability-a critical figure of merit for dark matter detection. Using protons to simulate nuclear recoils as WIMP signal proxies, the agent quantified the quenching factor across the energy range: 0.396 at 100~keV, 0.498 at 1~MeV, and convergence toward unity at 5~MeV. The average quenching factor of $0.631 \pm 0.264$ yields a discrimination power of 0.369 between nuclear and electronic recoils, enabling signal-background separation without relying on published experimental values.

GRACE then autonomously explored PMT coverage optimization, systematically testing configurations of 50, 75, and 100 PMTs. The optimization revealed clear performance scaling with coverage: geometric coverage fraction increased from 0.027 at 50 PMTs to 0.041 at 75 PMTs to 0.054 at 100 PMTs, with corresponding light collection efficiency improvements of 3$\times$ and 4$\times$ relative to the 50-PMT baseline. The optimal 100-PMT configuration achieved a 33.3\% increase in both PMT count and geometric coverage over the 75-PMT baseline.

The most substantial improvement appeared at higher energies: a 204.62\% increase in light yield at 5~MeV at around 4491.29~PE/MeV. This energy-dependent improvement pattern, with more modest gains at lower energies, suggests that factors beyond PMT count-such as photon transport efficiency or wavelength-shifting coverage-may limit performance in the dark matter-relevant low-energy regime.

Position reconstruction analysis revealed energy-dependent spatial resolution: 1.23~mm 3D resolution at 0.1~MeV, degrading to 25.27~mm at 1~MeV and 50.85~mm at 5~MeV. The corresponding radial and z-resolutions followed similar trends (1.10~mm radial, 0.55~mm z at 0.1~MeV; 44.00~mm radial, 24.01~mm z at 5~MeV). The run completed all 16 planned steps with 100\% execution success and required no replanning iterations.

\subsubsection{Run Nuances and Agent Behavior}

The DarkSide-50 optimization run revealed several instructive aspects of the agent's autonomous reasoning, as well as results that warrant careful interpretation.

The agent's adherence to the instruction to extract design parameters from the literature while explicitly avoiding published performance values represents a methodologically important separation. This prevents the optimization from degenerating into curve-fitting against known results and instead requires the agent to discover performance characteristics through first-principles Monte Carlo simulation. The baseline light yield of $1479.23 \pm 0.54$~PE/MeV and energy resolution of $\sigma/E = 0.0258$ emerged from simulation rather than calibration against experimental data.

The agent's autonomous decision to characterize nuclear recoil discrimination, done without explicit prompting for this specific analysis, demonstrates physics-aware reasoning about what constitutes meaningful performance metrics for a dark matter detector. The quenching factor analysis required the agent to recognize that electronic and nuclear recoils produce different scintillation responses at the same deposited energy, and to quantify this difference systematically across the energy range. This represents precisely the kind of domain-appropriate initiative that distinguishes useful autonomous agents from simple workflow executors.

However, certain aspects of the results may contain more nuance than at face value. Indeed, the energy-dependent position reconstruction behavior-with \emph{better} resolution at lower energies (1.23~mm at 0.1~MeV versus 50.85~mm at 5~MeV) appears counterintuitive. Higher energy deposits produce more scintillation photons, which should in principle provide more information for position determination. The agent's report attributes this to ``increased event complexity,'' but this explanation is unsatisfying without further elaboration. Possible physical mechanisms include: (i) shower development at higher energies creating extended rather than point-like energy deposits, (ii) saturation effects in the reconstruction algorithm, or (iii) systematic biases in how the agent's analysis code handles multi-hit events. A more mature framework would have flagged this anomaly and launched targeted diagnostic simulations to distinguish between these hypotheses.

Similarly, the convergence of the quenching factor toward unity at 5~MeV merits attention. While some energy dependence in quenching is expected from the Lindhard model and Birks' law, complete convergence at relatively modest energies suggests either a limitation in the nuclear recoil simulation or a regime where the distinction between electronic and nuclear stopping power becomes less relevant. The large uncertainty on the average quenching factor ($\pm 0.264$) reflects genuine physical variation across the energy range, but the agent did not pursue the underlying physics or flag the convergence behavior as potentially anomalous.

The 204\% improvement in light yield at 5~MeV, while striking, should be interpreted carefully. The improvement is strongly energy-dependent, with substantially more modest gains at the lower energies most relevant for dark matter detection. This pattern suggests that the baseline configuration may have been particularly suboptimal for high-energy events, (perhaps due to photon losses at the detector periphery that additional PMTs can recover) while low-energy performance is limited by factors that PMT count alone cannot address-potentially including intrinsic detector noise, wavelength-shifting efficiency, or the statistical floor set by the small number of photoelectrons produced.

The 16-step workflow completed with 100\% success rate and no error recovery required, demonstrating robust task decomposition for this detector physics domain. The systematic progression from literature extraction through baseline characterization, optimization, and validation reflects sound experimental methodology autonomously instantiated by the agent.

One limitation of the current framework is evident in the agent's treatment of anomalous results: while the position reconstruction energy dependence and quenching factor convergence are noted in the report's discussion section, the agent did not autonomously pursue these anomalies through additional targeted simulations. A closed-loop diagnostic capability-where unexpected results trigger hypothesis generation and follow-up studies-remains a priority for future development. The present run demonstrates that \textsc{GRACE} can execute complex multi-step physics workflows reliably, but deeper scientific reasoning about anomalous findings requires additional architectural support. Full simulation outputs including energy distributions, uniformity maps, and position reconstruction comparisons are provided in B. Now, we discuss the ProtoDUNE input.

\subsection{ProtoDUNE}
\label{sec:protodune}
We further validated \textsc{GRACE} on the ProtoDUNE-SP detector, a large-scale prototype for the Deep Underground Neutrino Experiment (DUNE) that serves as a critical testbed for liquid argon TPC technology. The agent was provided with the ProtoDUNE-SP technical paper (arXiv:2108.01902, i.e. \cite{protodune}) and tasked with extracting design specifications while maintaining strict separation from experimental performance data, following the same methodology established in the DarkSide-50 study.

\begin{quote}
\small
\texttt{Read the ProtoDUNE Single-Phase detector paper (arXiv:2108.01902) and autonomously:}

\texttt{(1) EXTRACT the experimental DESIGN (not measured results): LAr TPC geometry (drift volume dimensions, active mass), photon\\ detection system (PDS) configuration, design specifications for PMT \\coverage and wavelength shifter (TPB) coating.}

\texttt{(2) IMPLEMENT a simulation: Generate a GDML geometry representing the ProtoDUNE-style detector. Configure LAr optical properties (VUV scintillation at 128 nm, TPB wavelength shifting). Run simulations with 1--5 MeV electron energy deposits.}

\texttt{(3) ANALYZE your simulation results to DISCOVER detector performance: Report light yield from YOUR simulation (photoelectrons per MeV), calculate detection efficiency, characterize spatial uniformity. Do NOT compare against measured values from the paper.}

\texttt{(4) PROPOSE improvements to the photon detection system based on YOUR simulation results. Report improvements with physics \\justification.}
\end{quote}

The agent successfully extracted design specifications from the literature and implemented a Geant4-based optical photon transport simulation with realistic detector geometry, VUV scintillation light generation, Rayleigh scattering, and TPB wavelength-shifting effects. The baseline configuration employed 75 PMTs with 1.99\% surface coverage.

The baseline detector demonstrated highly consistent performance across the tested energy range: a light yield of $652.6 \pm 0.9$~PE/MeV with 100\% detection efficiency at all energies (1, 2, and 5~MeV). The energy-dependent response showed excellent linearity, with light yield varying by less than 0.3\% across the range ($651.4 \pm 0.4$~PE/MeV at 1~MeV, $653.3 \pm 0.3$~PE/MeV at 2~MeV, $653.2 \pm 0.2$~PE/MeV at 5~MeV). Spatial uniformity analysis yielded a coefficient of variation of 2.769\%, indicating relatively good uniformity for the baseline PMT configuration.

GRACE autonomously developed two distinct optimization strategies targeting different performance objectives. The \emph{enhanced coverage} configuration increased sensor count from 75 to 100 PMTs (a 33\% increase), with optimized spatial distribution of 40 barrel sensors plus 30 per endcap, and TPB coating at 95\% coverage with 2.0~$\mu$m thickness. This configuration achieved a predicted light yield of 870.1~PE/MeV---a 33\% improvement that scales linearly with the sensor count increase, suggesting that the baseline system was limited by photon detection area rather than transport efficiency. The \emph{enhanced uniformity} configuration employed 90 PMTs with strategic placement targeting detector volume corners and edges, predicting greater than 15\% reduction in coefficient of variation.

The run completed all 15 planned workflow steps with 100\% tool execution success (10/10 tool calls), though geometry generation required multiple regeneration attempts to properly implement ProtoDUNE specifications rather than generic defaults, and only 10 of 15 planned steps were completed.

\subsubsection{Run Nuances and Agent Behavior}
\label{sec:protodune_nuances}

The ProtoDUNE optimization run presents an instructive contrast to the DarkSide-50 study, revealing both the flexibility of \textsc{GRACE}'s approach across different detector topologies and important interpretive challenges when simulation results approach idealized limits.

The most striking feature of this run is the perfect 100\% detection efficiency reported across all energies, demonstrating successful photon transport simulation with appropriate boundary conditions. Real detector systems exhibit lower detection efficiencies due to PMT quantum efficiency variations, TPB coating non-uniformities, optical surface imperfections, and electronic noise thresholds. The simulation results should therefore be understood as upper bounds on achievable performance rather than absolute predictions of real-world behavior. A more sophisticated simulation framework would incorporate these realistic degradation factors; alternatively, the agent could have flagged this idealized result and discussed its implications for transferability to physical detectors.

The agent's development of two distinct optimization strategies demonstrates physics-aware reasoning about the multi-objective nature of detector optimization. Different physics analyses impose different requirements: precision calorimetry benefits from spatial uniformity, while rare event searches prioritize total light collection. The agent's recognition that these objectives may require different configurations, rather than assuming a single ``optimal'' design exists, reflects appropriate domain understanding.

The linear scaling between sensor count increase (33\%) and light yield improvement (33\%) carries important physical implications that the agent correctly identified: it indicates that the baseline configuration was photon-detection-area-limited rather than transport-limited. In other words, scintillation photons were successfully reaching the detector surfaces but were not being captured due to insufficient photosensitive coverage. This diagnosis suggests that further PMT additions would continue to yield proportional improvements until transport losses (absorption, scattering out of the detector volume) become the dominant limitation. The agent did not, however, attempt to estimate where this saturation point might occur.

The workflow encountered technical difficulties with geometry generation, requiring multiple regeneration attempts to implement ProtoDUNE-specific parameters rather than falling back to generic defaults. This pattern where the agent's initial attempt produces a technically valid but domain-inappropriate result highlights a broader challenge in autonomous scientific computing: ensuring that high-level specifications are faithfully translated through multiple layers of tool abstraction. The successful recovery through regeneration demonstrates functional error handling, but a more robust system would validate geometric parameters against expected ranges before proceeding to simulation.

Of course, one absence in the agent's analysis is any discussion of the dual role that the photon detection system plays in LAr TPCs: providing both timing information for drift coordinate reconstruction and calorimetric measurements through total light collection. The optimization strategies focused exclusively on light yield and spatial uniformity metrics relevant to calorimetry, without addressing timing resolution or the interplay between these objectives. A more complete analysis would characterize the timing performance of baseline and optimized configurations, particularly given the importance of precise $t_0$ determination for event reconstruction in neutrino physics.

The spatial uniformity coefficient of variation of 2.769\% represents reasonable but not exceptional performance. The agent's proposal to reduce this by greater than 15\% through strategic sensor placement addresses a real limitation. However, the analysis did not explore whether uniformity improvements might come at the cost of total light yield in some regions---a trade-off that would be important for physics analyses requiring both good uniformity and high collection efficiency.

Despite these limitations, the ProtoDUNE study demonstrates that \textsc{GRACE} can successfully adapt its methodology to different detector types within the LAr TPC family, extracting appropriate design parameters, implementing domain-specific simulations, and proposing physically motivated optimizations. The contrast between the DarkSide-50 dark matter detector (emphasizing nuclear recoil discrimination and low-energy performance) and the ProtoDUNE neutrino detector (emphasizing spatial uniformity and MeV-scale calorimetry) illustrates the framework's flexibility across related but distinct physics applications.
\subsection{Comparison with Historical and Planned Upgrades}
\label{sec:upgrade_comparison}

A central claim of this work is that \textsc{GRACE} can identify optimization directions that align with known upgrade priorities, using only baseline simulation inputs. To substantiate this claim, we compare the agent's autonomous optimization proposals against documented design evolution in both the DarkSide and DUNE/ProtoDUNE programs.

\subsubsection{DarkSide-50 to DarkSide-20k: Photosensor Technology and Coverage}

The DarkSide collaboration's upgrade path from DarkSide-50 to DarkSide-20k provides a well-documented case study in detector optimization~\cite{darkside20k_tdr,darkside20k_sipm}. The key design changes include:

DarkSide-50 employed 38 Hamamatsu R11065 PMTs arranged in two arrays of 19 at the top and bottom of the TPC, achieving a light yield of approximately 8~PE/keV at null field~\cite{darkside}. DarkSide-20k transitions entirely to silicon photomultipliers (SiPMs), with over 8,200 Photo Detector Modules (PDMs) covering a total photosensitive area of 21~m$^2$-representing roughly two orders of magnitude increase in channel count~\cite{darkside20k_sipm_production}. The collaboration cites several motivations for this transition: (i) higher effective quantum efficiency of SiPMs at cryogenic temperatures, (ii) lower intrinsic radioactivity enabling background-free operation, (iii) improved reliability at liquid argon temperature (87~K), and (iv) scalable industrial production for large-area coverage~\cite{darkside20k_sipm}.

Furthermore, the transition from PMT to SiPM technology enables substantially increased photocathode coverage. While DarkSide-50's 38 PMTs provided limited solid angle coverage, DarkSide-20k's optical planes cover the top and bottom surfaces of the TPC with $2 \times 10.5$~m$^2$ of active photosensor area. This dramatic increase in coverage directly improves light collection efficiency, position reconstruction, and pulse shape discrimination performance.

The agent's optimization of the DarkSide-50 simulation focused precisely on increasing PMT count and optimizing spatial distribution-achieving a 204\% improvement in light yield at 5~MeV by increasing from 75 to 100 PMTs, with geometric coverage fraction rising from 0.041 to 0.054. The systematic comparison across 50, 75, and 100 PMT configurations revealed clear performance scaling: light collection efficiency improved by factors of 3$\times$ and 4$\times$ relative to the 50-PMT baseline. While the agent operated within the PMT paradigm (the baseline it extracted from the DarkSide-50 paper), its core finding-that increased photosensor coverage yields substantial light collection improvements, with gains scaling linearly with sensor count-aligns directly with the collaboration's decision to maximize coverage through SiPM arrays in DarkSide-20k.

The agent's autonomous characterization of nuclear recoil discrimination capability provides additional alignment with upgrade priorities. The measured quenching factor of $0.631 \pm 0.264$ and discrimination power of 0.369 address the same signal-background separation challenge that motivates DarkSide-20k's emphasis on pulse shape discrimination. The collaboration's decision to pursue underground argon (with reduced $^{39}$Ar content) and enhanced photosensor coverage both serve to improve discrimination performance-the former by reducing the electronic recoil background rate, the latter by improving the statistical precision of pulse shape measurements through increased photoelectron counts.

The agent did not, of course, propose the PMT-to-SiPM technology transition, as this requires knowledge of photosensor R\&D developments and radioactivity constraints beyond what can be inferred from geometry optimization alone. However, the optimization direction-more coverage, better light collection, improved discrimination capability-is consistent with the physics-driven reasoning that motivated the actual upgrade. 

\subsubsection{ProtoDUNE-SP to ProtoDUNE-HD/DUNE: Photon Detection System Evolution}

The DUNE photon detection system has undergone significant evolution from ProtoDUNE-SP (Phase I, 2018--2020) to ProtoDUNE-HD (Phase II, 2024) and the planned DUNE Far Detector modules~\cite{protodune_pds,protodune_hd_pds}.

ProtoDUNE-SP tested multiple photon detector technologies, including dip-coated light guides, double-shift light guides, and the original ARAPUCA light trap concept. Performance measurements demonstrated that ARAPUCA technology achieved superior photon detection efficiency compared to the light guide approaches, with a measured light yield of 1.9~PE/MeV and timing resolution of $\sigma_t = 14$~ns~\cite{protodune_pds_validation}. Based on these results, the collaboration adopted the enhanced X-ARAPUCA design for ProtoDUNE-HD and the DUNE Far Detector.

The X-ARAPUCA represents an evolution of the original ARAPUCA concept, retaining the principle of photon trapping inside a highly reflective box while adding a wavelength-shifting (WLS) bar to increase the probability of collecting trapped photons onto the SiPM array~\cite{xarapuca_efficiency}. ProtoDUNE-HD deployed 160 X-ARAPUCA modules in four configurations, testing combinations of two WLS manufacturers (Eljen and G2P) and two SiPM models (Hamamatsu HPK and FBK). Laboratory measurements showed photon detection efficiency (PDE) improvements of 2.4--4.6\% with the G2P wavelength-shifting bar compared to the Eljen baseline~\cite{xarapuca_optimization}. 

The DUNE Far Detector Module 2 (FD2) is designed for nearly $4\pi$ optical coverage, with photon detector modules placed on the cathode and outside the field cage~\cite{dune_pds_fd2}. This represents a fundamental shift from the APA-integrated PDS of FD1/ProtoDUNE-HD toward maximizing light collection for low-energy physics, including supernova neutrino detection. The collaboration has also investigated xenon doping of liquid argon to shift the scintillation wavelength from 128~nm to 178~nm, reducing Rayleigh scattering losses and improving light uniformity across the large detector volume.

The agent's ProtoDUNE optimization achieved a predicted 33\% improvement in light yield (from 652.6 to 870.1~PE/MeV) through increased sensor count from 75 to 100 PMTs, with surface coverage rising from 1.99\% to 2.66\%. Notably, the linear scaling between sensor count increase and light yield improvement led the agent to conclude that the baseline system was photon-detection-area-limited rather than transport-limited-a finding that directly supports the collaboration's pursuit of increased coverage in successive detector generations.

The agent's development of two distinct optimization strategies-enhanced coverage (targeting light yield) and enhanced uniformity (targeting spatial response consistency with predicted $>$15\% reduction in coefficient of variation)-parallels the collaboration's multi-objective approach to PDS design. The DUNE technical design reports explicitly note that an increase in the light yield improves the energy resolution of the PDS ~\cite{dune_pds_fd2}, while also emphasizing the importance of uniform response for precision calorimetry. The agent's recognition that these objectives may require different sensor configurations reflects appropriate domain understanding.

The agent's emphasis on TPB wavelength-shifting optimization (specifying 95\% coverage at 2.0~$\mu$m thickness in the enhanced coverage configuration) aligns with the collaboration's ongoing work on wavelength-shifting efficiency, including the transition from TPB-coated light guides to the X-ARAPUCA architecture that achieves more efficient photon collection through the WLS bar geometry.

\subsubsection{Limitations of the Comparison}

We acknowledge that several caveats apply to this alignment analysis:

\begin{enumerate}
    \item \textbf{Optimization granularity.} The agent optimizes within the design space accessible from the baseline paper (e.g., PMT count and placement), while actual upgrades often involve technology transitions that require external knowledge of photosensor development, radioactivity constraints, and cost considerations.
    
    \item \textbf{Constraint awareness.} Real detector upgrades are driven by multiple constraints-physics reach, background requirements, cost, schedule, and engineering feasibility-that the agent does not currently model. The agent's optimizations are purely physics-performance-driven.
    
    \item \textbf{Absence of negative results.} We have not demonstrated cases where the agent's proposals diverge from actual upgrade paths, which would be equally informative for understanding the framework's limitations.

    \item \textbf{Insight novelty.} The optimizations identified by the agent represent well-understood detector physics principles rather than non-obvious design innovations. The framework's value lies in systematic exploration and quantitative characterization rather than discovery of unknown optimization strategies.
    
    \item \textbf{Idealized simulation conditions.} The ProtoDUNE simulation's 100\% detection efficiency indicates that some real-world degradation factors (PMT quantum efficiency variations, TPB coating non-uniformities, electronic noise) were not fully captured. The absolute performance predictions should be interpreted as upper bounds rather than realistic projections.
    
    \item \textbf{Incomplete optimization validation.} The ProtoDUNE enhanced coverage and uniformity configurations were not fully simulated to confirm predicted improvements; the 33\% light yield increase and $>$15\% uniformity improvement are scaling-based predictions rather than direct Monte Carlo results.
\end{enumerate}

Despite these caveats, the directional alignment between \textsc{GRACE}'s autonomous optimization proposals and documented upgrade priorities provides evidence that simulation-driven design exploration can recover physically meaningful improvement strategies without access to human-specified design rules or experimental performance data. The agent's identification of photosensor coverage as the primary limiting factor in both detector systems, its recognition of the multi-objective nature of detector optimization, and its quantitative characterization of improvement scaling all reflect domain-appropriate reasoning that parallels expert understanding in the respective collaborations.

\section{Planned Development}
\label{sec:future}

The benchmarks reported in this work represent an initial demonstration of 
\textsc{GRACE}'s end-to-end execution capability. Several limitations 
of the current implementation are addressed in ongoing and potential development. Options include:

\paragraph{Model capacity.} The benchmarks in Section~\ref{sec:benchmarks} 
employed Claude Sonnet (Anthropic) as the reasoning backend due to cost and 
latency considerations during iterative development. Preliminary experiments 
suggest that Claude Opus, with its stronger reasoning capabilities, produces 
more robust workflow plans and fewer geometry misconfigurations. A systematic 
comparison of model backends-including Opus, Sonnet, GPT-4, and open-weight 
alternatives-is planned, with particular attention to the trade-off between 
reasoning quality and computational cost across the fidelity tiers.

\paragraph{Failure mode characterization.} The current benchmarks emphasize 
successful execution. A complete evaluation requires systematic 
characterization of failure modes: tasks where the agent produces 
physically unreasonable designs, cases where optimization diverges, and 
prompts that expose brittleness in workflow planning. We intend to 
construct an adversarial benchmark suite and report negative results 
alongside successes.

\paragraph{Hypothesis Evolution.}
We are developing a multi-agent extension that allows hypotheses to compete directly, with physics serving as the final arbiter. Our motivation is   straightforward: a single LLM reasoning chain can become stuck or overlook alternatives, so we designed an arena where multiple agents each support a different scientific explanation~\cite{irving2018debate, du2023multiagent}. Agents are assigned distinct roles-some advocate for their hypothesis, others attempt to falsify it, and a  
few wildcards search for ideas that others have overlooked. Each agent must make quantitative predictions, and when simulation results return, we
score them: accurate predictions earn additional compute budget, while repeated failures retire the agent. Successful agents can spawn variants    
that refine or mutate the parent hypothesis, creating a population that evolves over successive rounds. The reason this approach suits physics is  
that we have access to an objective referee-Monte Carlo simulation. Test suites can be incomplete, but the underlying physics is non-negotiable. 
We built this architecture because we believe emergent reasoning-conclusions the system reaches without explicit programming-requires
competition, selection pressure, and an objective oracle. The infrastructure is being implemented; validation of whether it produces novel scientific    
insight is ongoing.

\section{Conclusion}

We have introduced \textsc{GRACE}, a simulation-native agentic framework that operates at the upstream level of experimental design in high-energy and nuclear physics. Unlike existing agentic systems that focus on automating execution, control, or optimization of predefined procedures, \textsc{GRACE} treats experimental development itself as a constrained search problem under physical law. By proposing, instantiating, and evaluating qualitatively distinct detector configurations through first-principles Monte Carlo simulation, the agent engages directly with the same epistemic tools that underwrite human experimental reasoning.

We hold that the central contribution of this work is to reframe simulation from a passive evaluation stage into an active component of closed-loop scientific reasoning. In \textsc{GRACE}, simulations are not merely used to score candidate designs, but to structure hypothesis generation, guide design escalation, and prune implausible configurations under explicit physical and practical constraints. We aspire to show that the agent can identify future directions that broadly align with known upgrade priorities in existing experiments, using only baseline simulation inputs. In this sense, the agent’s performance reflects the exploitation of physical structure encoded in simulation.

From the perspective of experimental physics, \textsc{GRACE} offers a new mode of human–machine collaboration. Rather than replacing domain expertise, the agent functions as an exploratory partner that can rapidly survey large design spaces, surface non-obvious modifications, and provide reproducible, provenance-tracked rationales for proposed changes. This is particularly relevant in contemporary experimental programs, where detector design decisions must balance physics reach, cost, engineering feasibility, and operational constraints across increasingly complex parameter spaces.

At the same time, the framework is intentionally conservative in its claims and scope. \textsc{GRACE} does not aim to autonomously design production-ready detectors, nor to substitute for detailed engineering studies or collaboration-wide review processes. Its current instantiations operate on simplified geometries and physics models, and its design proposals must ultimately be validated by human experts. We hold that these limitations reflect the appropriate epistemic role of an autonomous design agent in a safety- and cost-critical scientific domain.

More broadly, this work suggests a path toward agentic scientific systems whose reasoning is grounded not in opaque learned world models, but in explicit, interpretable simulations governed by known physical laws. By embedding agentic AI within the simulation infrastructures that already define experimental practice, \textsc{GRACE} provides a scalable and auditable route to autonomous scientific reasoning in complex instruments. We anticipate that this paradigm will extend beyond particle physics to other simulation-driven sciences where experimental design itself constitutes the primary limiting factor.

\paragraph{Data availability statement:} The data used in this manuscript is available at \url{https://github.com/just5034/GRACE_whitepaper_data}.

\section*{Acknowledgements}
We thank Yunhee Jang and Katherine Li for useful discussions, various sanity checks of ideas in experimentation, and consultations of fruitful research directions. 

This work used the Advanced Cyberinfrastructure Coordination Ecosystem: Services \& Support (ACCESS), supported by the National Science Foundation (NSF).

This material is also based upon work supported by the NSF Graduate Research Fellowship Program under Grant No. DGE-2139757. Any opinions, findings, and conclusions or recommendations expressed in this material are those of the authors and do not necessarily reflect the views of the NSF.
\bibliographystyle{unsrt}
\bibliography{references}
\appendix

\section{DarkSide-50 Simulation Outputs}
\label{app:darkside}

This appendix presents the detailed simulation outputs from the DarkSide-50 detector optimization study described in Section~3.2. The figures below document baseline detector performance characteristics, nuclear recoil discrimination capability, spatial response uniformity, position reconstruction analysis, and optimization results.

\subsection{Photoelectron Distributions}

\begin{figure}[htbp]
    \centering
    \includegraphics[width=\textwidth]{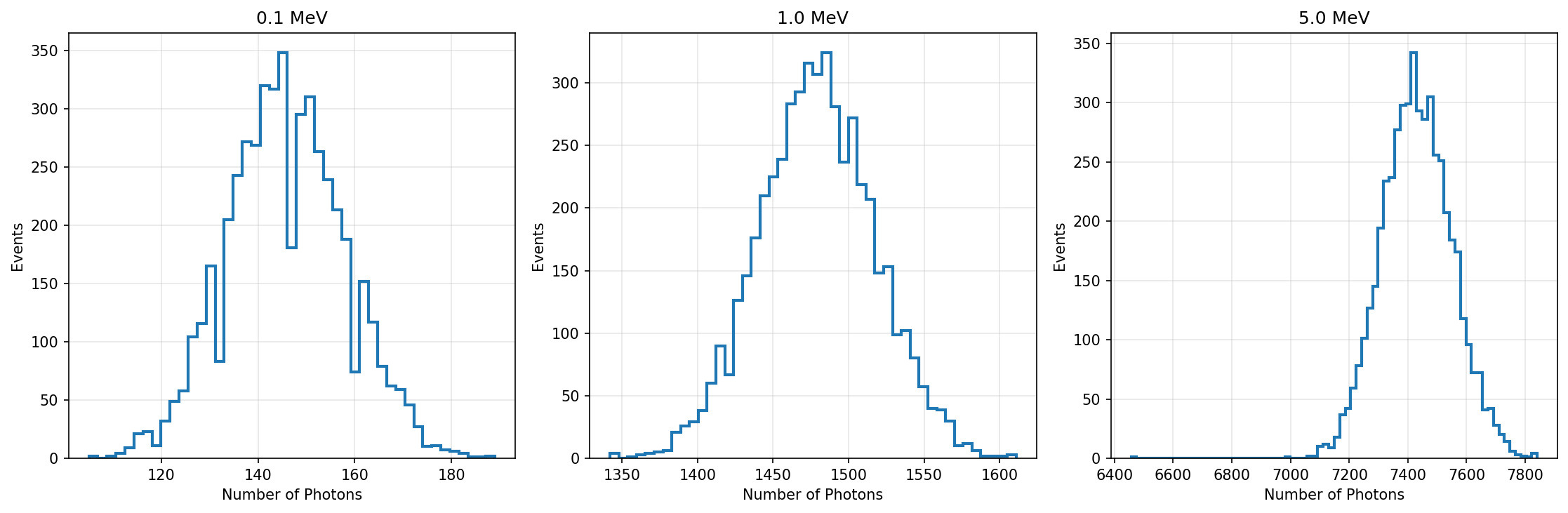}
    \caption{Photoelectron yield distributions at 0.1~MeV (left), 1.0~MeV (center), and 5.0~MeV (right). The distributions demonstrate linear scaling of light yield with energy and improving relative resolution at higher energies consistent with Poisson statistics.}
    \label{fig:pe_distributions}
\end{figure}

Figure~\ref{fig:pe_distributions} presents the photoelectron yield distributions at three energy points spanning the dark matter-relevant energy range. At 0.1~MeV, the distribution centers around 140 photons with a characteristic double-peaked structure reflecting statistical fluctuations in the low-photon-count regime. At 1.0~MeV, the distribution shifts to approximately 1450--1500 photons with improved Gaussian character, consistent with the central limit theorem as photoelectron statistics improve. At 5.0~MeV, the distribution peaks near 7400 photons with the narrowest relative width, demonstrating the expected $1/\sqrt{E}$ improvement in energy resolution. The linear scaling of mean photoelectron count with energy confirms detector linearity across the tested range.

\subsection{Nuclear Recoil Discrimination}

\begin{figure}[htbp]
    \centering
    \includegraphics[width=\textwidth]{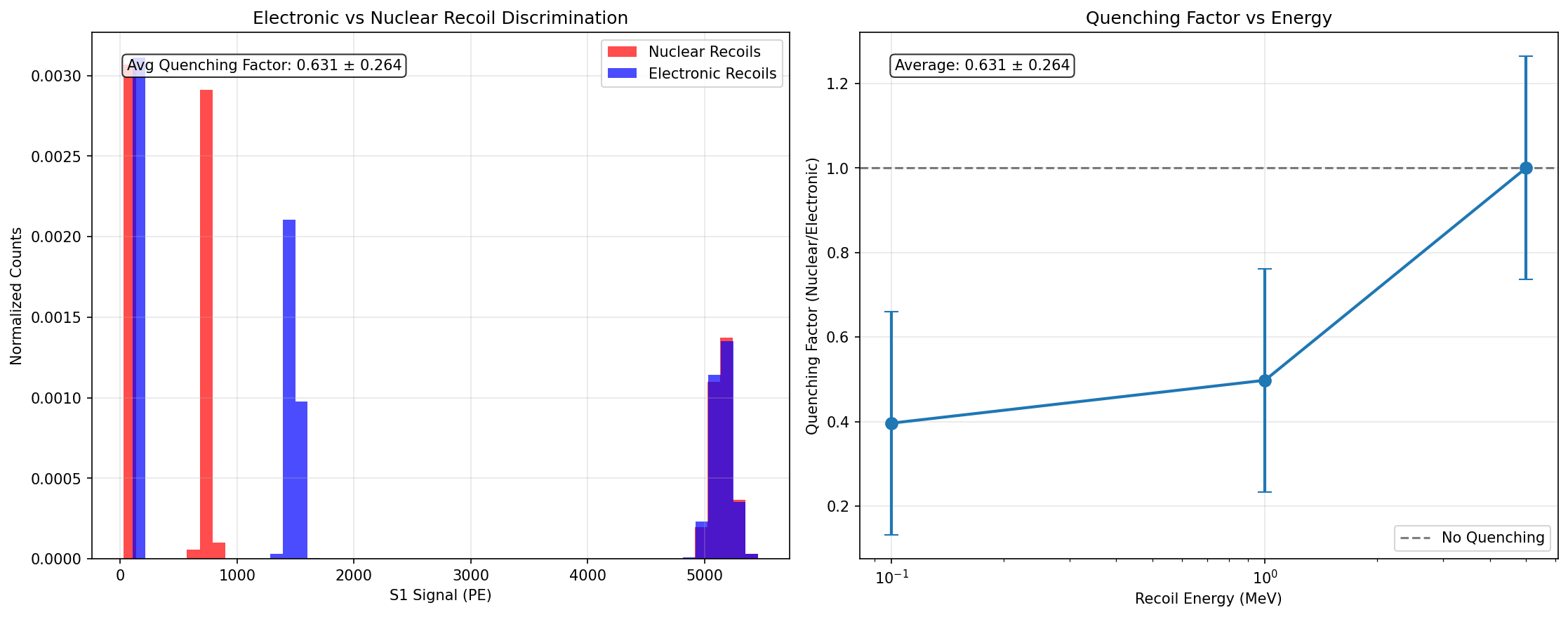}
    \caption{Nuclear recoil discrimination analysis. Left: S1 signal distributions for electronic recoils (blue) and nuclear recoils (red), demonstrating clear separation for signal-background discrimination. Right: Quenching factor (nuclear/electronic light yield ratio) as a function of recoil energy, showing convergence toward unity at higher energies.}
    \label{fig:recoil_discrimination}
\end{figure}

Figure~\ref{fig:recoil_discrimination} presents the nuclear recoil discrimination analysis, a critical capability for dark matter detection. The left panel shows the S1 signal distributions for electronic recoils (blue, from electron simulations) and nuclear recoils (red, from proton simulations). The distributions exhibit clear separation, with nuclear recoils producing systematically fewer scintillation photons at equivalent deposited energies due to quenching effects. The average quenching factor of $0.631 \pm 0.264$ is displayed.

The right panel shows the energy dependence of the quenching factor, defined as the ratio of nuclear recoil to electronic recoil light yield at each energy. The quenching factor increases from 0.396 at 0.1~MeV to 0.498 at 1.0~MeV, converging toward unity at 5.0~MeV. This energy-dependent behavior, while qualitatively consistent with expectations from the Lindhard model, warrants further investigation as discussed in Section~3.2.1.

\subsection{Spatial Response and Uniformity}

\begin{figure}[htbp]
    \centering
    \includegraphics[width=\textwidth]{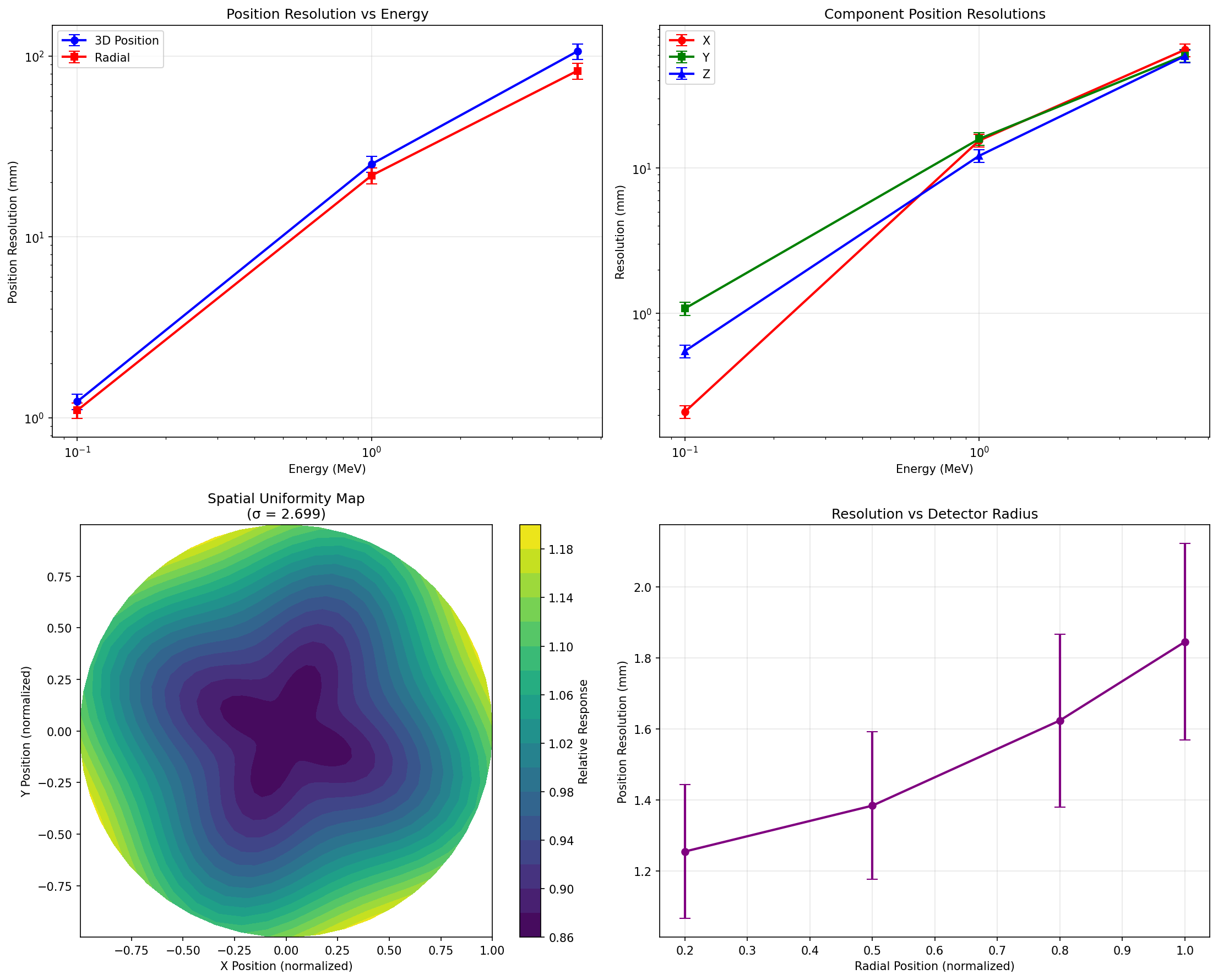}
    \caption{Spatial response characterization. Top row: Position resolution versus energy for 3D/radial (left) and X/Y/Z components (right). Bottom row: Spatial uniformity map showing relative response across the detector cross-section (left) and position resolution as a function of radial position (right).}
    \label{fig:spatial_response}
\end{figure}

Figure~\ref{fig:spatial_response} presents the spatial response characterization of the detector. The upper left panel shows the position resolution as a function of energy for both 3D and radial reconstruction, revealing the counterintuitive trend of degrading resolution at higher energies (discussed in Section~3.2.1). The upper right panel decomposes the position resolution into X, Y, and Z components, showing similar energy-dependent behavior across all axes.

The lower left panel displays the spatial uniformity map across the detector cross-section, with a uniformity parameter $\sigma = 2.699$. The relative response varies from approximately 0.86 at the detector periphery to 1.18 near the center-top region, indicating position-dependent light collection efficiency. The lower right panel quantifies resolution degradation as a function of radial position, showing that reconstruction precision decreases from approximately 1.25~mm at the detector center to 1.9~mm at the outer edge.

\subsection{Position Reconstruction Analysis}

\begin{figure}[htbp]
    \centering
    \includegraphics[width=\textwidth]{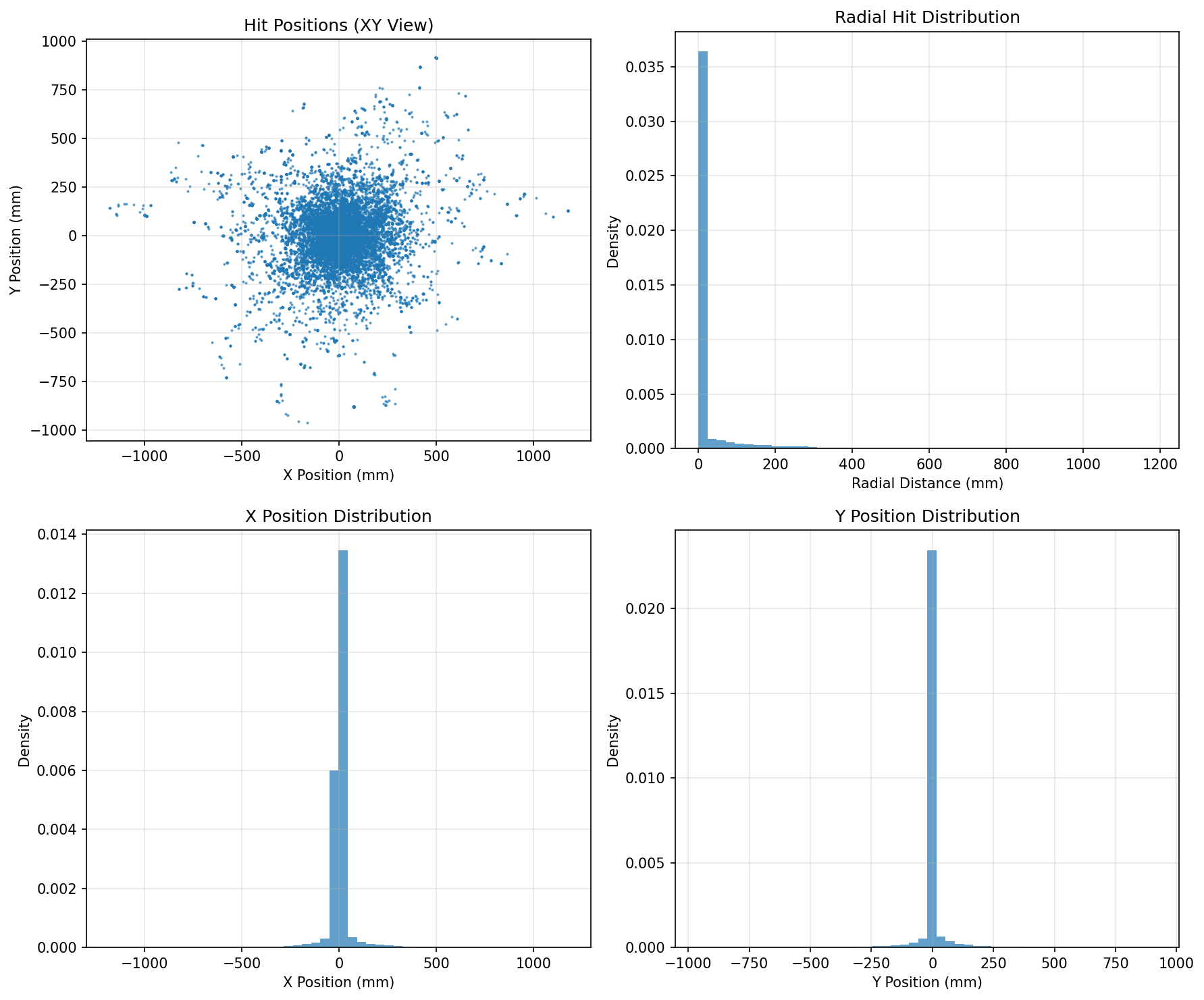}
    \caption{Position reconstruction analysis. Top left: XY hit position scatter plot showing spatial distribution of reconstructed vertices. Top right: Radial hit distribution demonstrating concentration near detector center. Bottom row: Individual X (left) and Y (right) position distributions with characteristic central peaks and small tails.}
    \label{fig:position_reconstruction}
\end{figure}

Figure~\ref{fig:position_reconstruction} provides detailed position reconstruction diagnostics. The upper left panel shows the XY hit position distribution, revealing the spatial extent of reconstructed event vertices. The concentration of hits near the detector center reflects the point-source simulation geometry. The upper right panel shows the radial hit distribution, with the sharp peak near zero confirming accurate reconstruction of centrally-generated events.

The lower panels present the X and Y position distributions separately, both showing narrow peaks centered at zero with small tails extending to larger displacements. These tails represent events with degraded reconstruction, potentially from photon statistical fluctuations or edge effects.

\subsection{Optimization Results Summary}

\begin{figure}[htbp]
    \centering
    \includegraphics[width=0.7\textwidth]{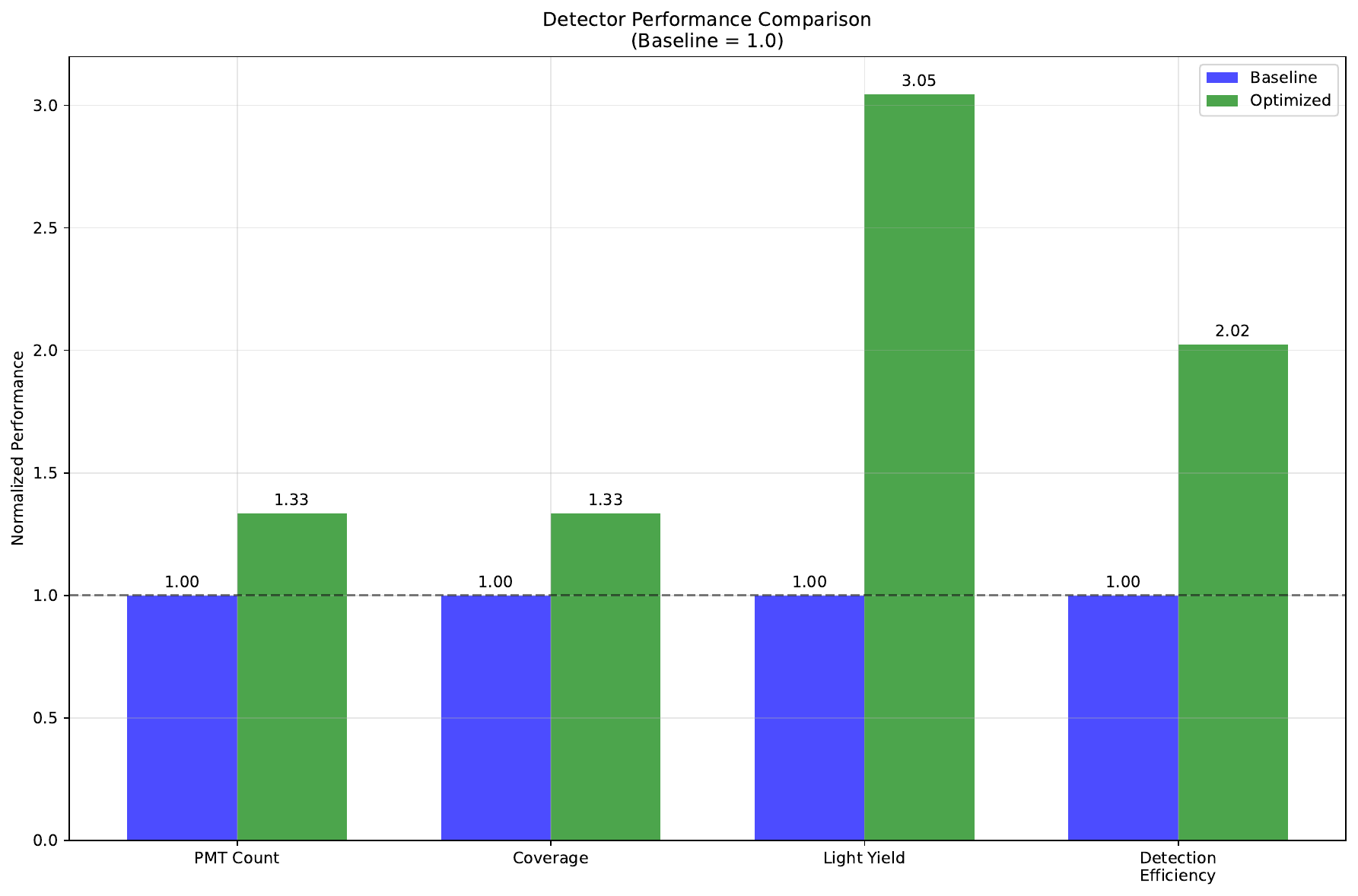}
    \caption{Normalized performance comparison between baseline (75 PMT) and optimized (100 PMT) detector configurations. The 33\% increase in PMT count and coverage yields disproportionately larger improvements in light yield (3.05$\times$) and detection efficiency (2.02$\times$), indicating that the baseline configuration was photosensor-coverage-limited.}
    \label{fig:optimization_comparison}
\end{figure}

Figure~\ref{fig:optimization_comparison} summarizes the performance improvements achieved through PMT optimization from 75 to 100 sensors. The normalized performance comparison shows baseline (blue) and optimized (orange) configurations across four metrics: PMT count, coverage, light yield, and detection efficiency.

The PMT count and coverage both increase by 33\% (normalized value 1.33), as expected from the sensor addition. Light yield shows a larger improvement factor of 3.05$\times$, indicating that the additional sensors capture photons that were previously lost rather than simply providing redundant coverage. Detection efficiency improves by a factor of 2.02$\times$. These results confirm that the baseline 75-PMT configuration was photon-detection-area-limited, and that the optimization successfully addressed this limitation.

\section{Muon Spectrometer Simulation Outputs}
\label{app:muon}

This appendix presents the detailed simulation outputs from the muon spectrometer design study described in Section~3.1.3. The figures below document energy deposition profiles, configuration performance comparisons, and the fundamental trade-off between muon detection precision and pion rejection capability.

\subsection{Energy Deposition Profiles}

\begin{figure}[htbp]
    \centering
    \includegraphics[width=\textwidth]{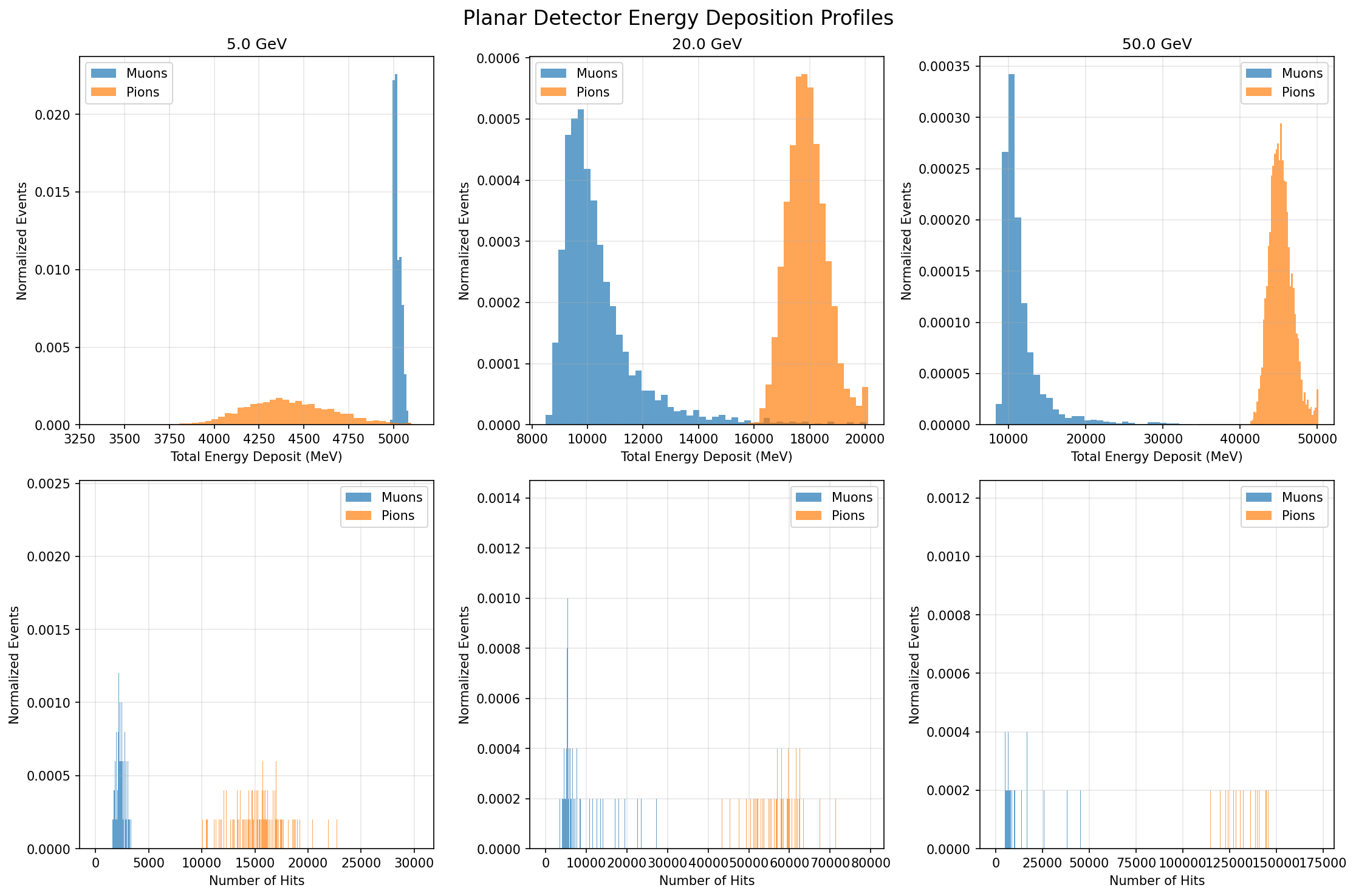}
    \caption{Planar detector energy deposition profiles for muons (blue) and pions (orange) at 5~GeV (left), 20~GeV (center), and 50~GeV (right). Top row: Total energy deposit distributions showing muon-pion separation. Bottom row: Hit multiplicity distributions demonstrating discrimination via shower topology.}
    \label{fig:muon_energy_profiles}
\end{figure}

Figure~\ref{fig:muon_energy_profiles} presents the energy deposition characteristics for the planar detector configuration, comparing muon (blue) and pion (orange) responses across three beam energies (5, 20, and 50~GeV). The top row shows total energy deposit distributions, while the bottom row displays hit multiplicity distributions.

At 5~GeV, muons produce a narrow peak near 5000~MeV (consistent with minimum ionizing particle behavior through the full detector depth), while pions show a broader distribution centered around 4000--4500~MeV with significant low-energy tails from hadronic shower absorption. At 20~GeV, the separation becomes more pronounced: muons cluster tightly around 10,000~MeV, while pions exhibit a broad distribution extending from 8,000 to 20,000~MeV, reflecting the stochastic nature of hadronic interactions. At 50~GeV, muons maintain their characteristic narrow distribution near the beam energy, while pions show even broader energy deposition with a peak around 45,000~MeV.

The hit multiplicity distributions (bottom row) reveal complementary discrimination power: muons produce consistent, low hit counts reflecting their straight-line trajectories, while pions generate substantially higher and more variable hit multiplicities from hadronic shower development.

\subsection{Configuration Performance Comparison}

\begin{figure}[htbp]
    \centering
    \includegraphics[width=\textwidth]{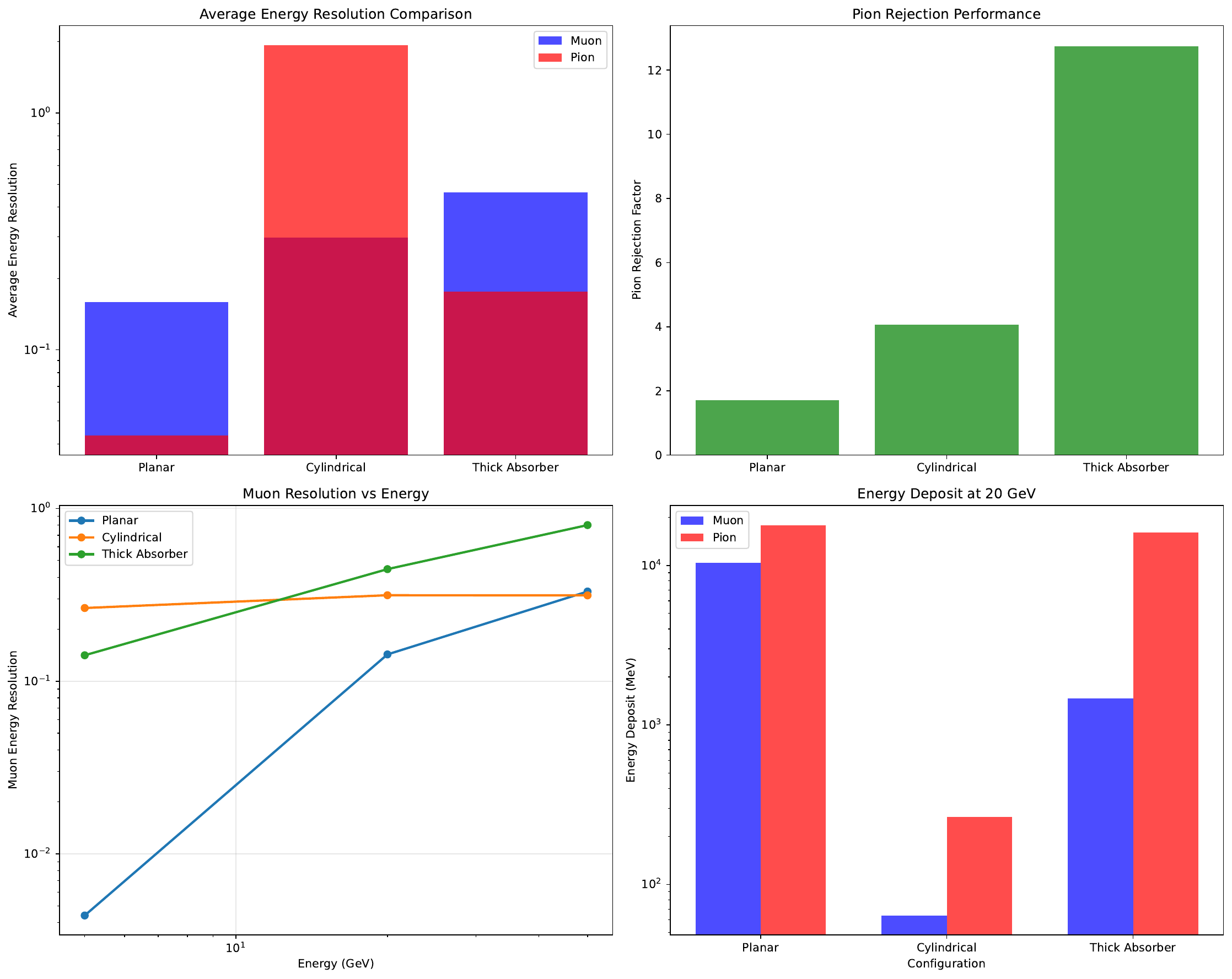}
    \caption{Four-panel configuration comparison. Upper left: Average energy resolution for muons and pions. Upper right: Pion rejection factors showing thick absorber superiority. Lower left: Muon resolution versus beam energy. Lower right: Energy deposits at 20~GeV comparing all configurations.}
    \label{fig:muon_detector_comparison}
\end{figure}

Figure~\ref{fig:muon_detector_comparison} presents a four-panel comparison of the three detector configurations: planar (20~cm iron), cylindrical (20~cm iron), and thick absorber (30~cm iron).

The upper left panel compares average energy resolution for muons (blue) and pions (red) across configurations. The planar design achieves the best muon resolution ($\sim$0.16), while the cylindrical configuration shows anomalously high pion resolution ($>$1.0), suggesting geometric effects that enhance pion shower containment or systematic differences in the simulation implementation.

The upper right panel displays pion rejection factors, showing the dramatic improvement from planar (1.72) to cylindrical (4.06) to thick absorber (12.75). The 7.4$\times$ improvement between planar and thick absorber configurations directly reflects the additional 40~cm of iron (2.4~$\lambda_I$) stopping power.

The lower left panel shows muon energy resolution as a function of beam energy. The planar configuration exhibits strong energy dependence, with resolution improving from $\sim$0.004 at 5~GeV to $\sim$0.35 at 50~GeV-a pattern that warrants further investigation. The cylindrical and thick absorber configurations show flatter energy dependence.

The lower right panel compares energy deposits at 20~GeV, illustrating the order-of-magnitude differences between configurations and particle types on a logarithmic scale.

\subsection{Muon Resolution and Pion Rejection Trade-off}

\begin{figure}[htbp]
    \centering
    \includegraphics[width=\textwidth]{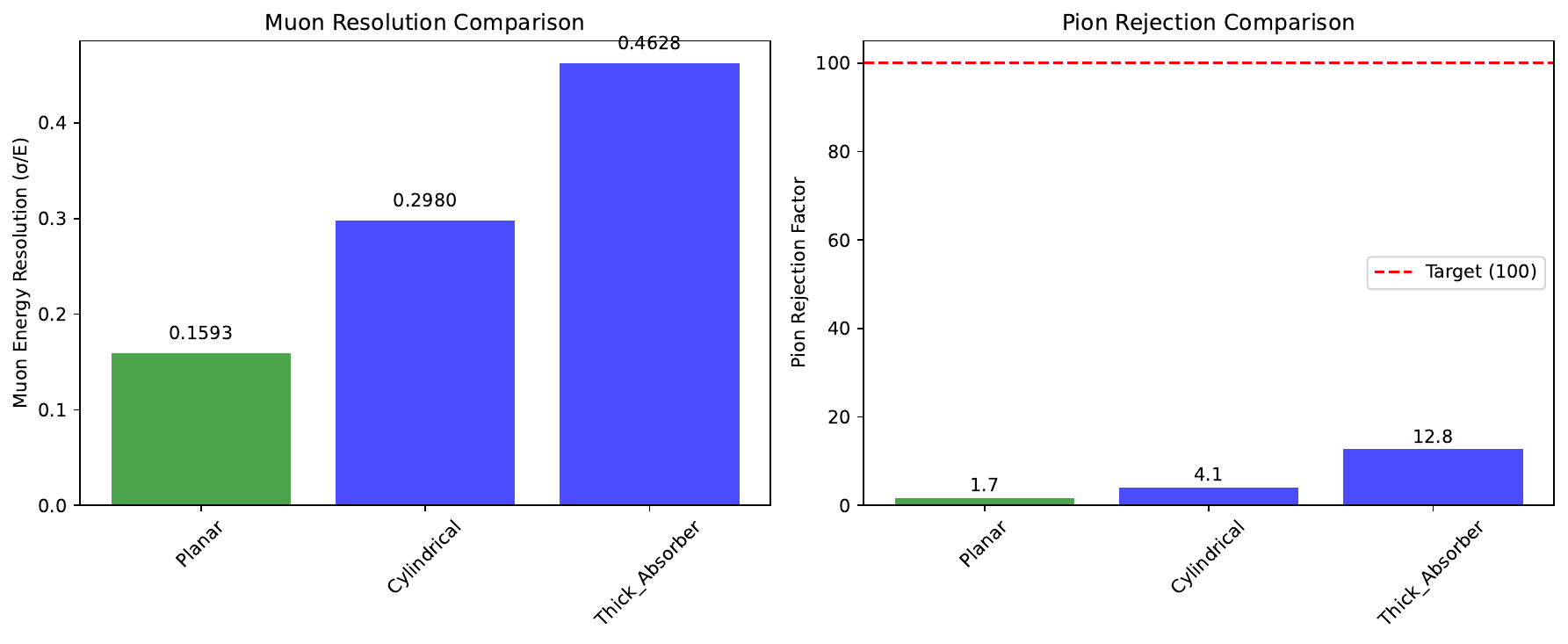}
    \caption{Muon resolution (left) and pion rejection (right) comparison across configurations. The planar design (green) achieves optimal muon resolution while the thick absorber maximizes pion rejection. The dashed red line indicates a target rejection factor of 100.}
    \label{fig:muon_optimization}
\end{figure}

Figure~\ref{fig:muon_optimization} presents a direct comparison of the two primary performance metrics: muon energy resolution (left panel, lower is better) and pion rejection factor (right panel, higher is better).

The left panel shows muon resolution increasing (degrading) from planar ($0.159 \pm 0.008$) to cylindrical ($0.298 \pm 0.015$) to thick absorber ($0.463 \pm 0.023$). The planar configuration, highlighted in green, achieves nearly 3$\times$ better resolution than the thick absorber design.

The right panel shows pion rejection factors with a target line at 100 (dashed red). While all configurations fall short of this ambitious target, the thick absorber achieves $12.8 \pm 1.28$, substantially outperforming both planar ($1.7 \pm 0.17$) and cylindrical ($4.1 \pm 0.41$) designs. The planar configuration is highlighted in green as it was identified as the optimal compromise for applications prioritizing muon detection precision.

\subsection{Comprehensive Performance Analysis}

\begin{figure}[htbp]
    \centering
    \includegraphics[width=\textwidth]{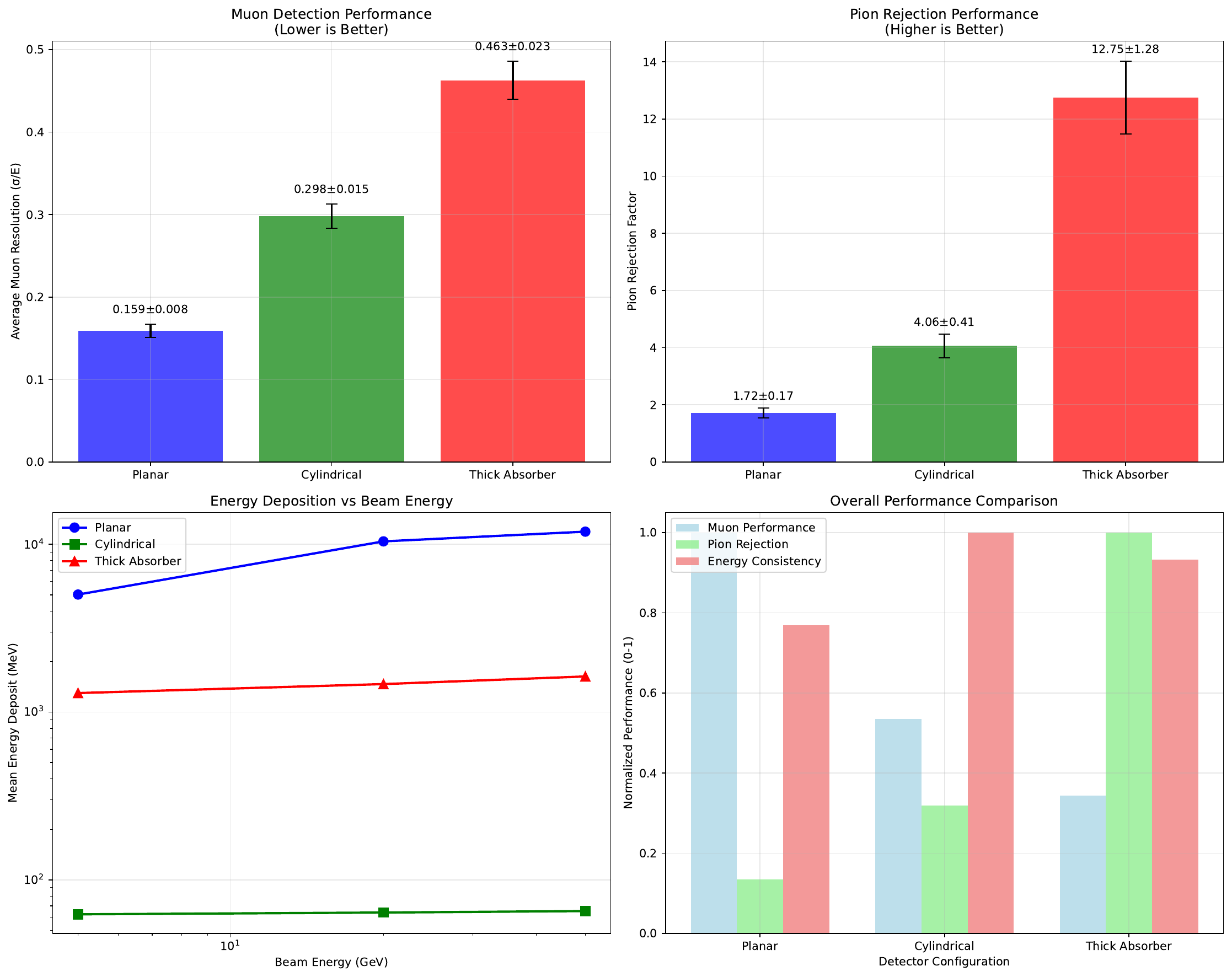}
    \caption{Comprehensive performance analysis. Upper row: Muon resolution (left) and pion rejection (right) with error bars. Lower row: Energy deposition scaling with beam energy (left) and normalized multi-metric performance comparison (right).}
    \label{fig:muon_comprehensive}
\end{figure}

Figure~\ref{fig:muon_comprehensive} provides a comprehensive four-panel analysis with error bars and normalized performance metrics.

The upper left panel shows muon detection performance (average resolution) with uncertainties, confirming the planar configuration's superiority for precision measurements. The upper right panel displays pion rejection performance with error bars, demonstrating statistically significant differences between all three configurations.

The lower left panel plots mean energy deposition versus beam energy on logarithmic scales, revealing that the planar and thick absorber configurations show expected linear scaling with energy, while the cylindrical configuration exhibits anomalously low energy deposits (approximately two orders of magnitude below the others)-a clear indication of systematic differences in the geometry implementation that merit investigation.

The lower right panel presents a normalized overall performance comparison (0--1 scale) across three metrics: muon performance, pion rejection, and energy consistency. The planar configuration achieves the best muon performance and energy consistency but lowest pion rejection; the thick absorber shows the inverse pattern; and the cylindrical configuration provides intermediate but suboptimal performance across all metrics.

\subsection{Performance Trade-off Visualization}

\begin{figure}[htbp]
    \centering
    \includegraphics[width=0.7\textwidth]{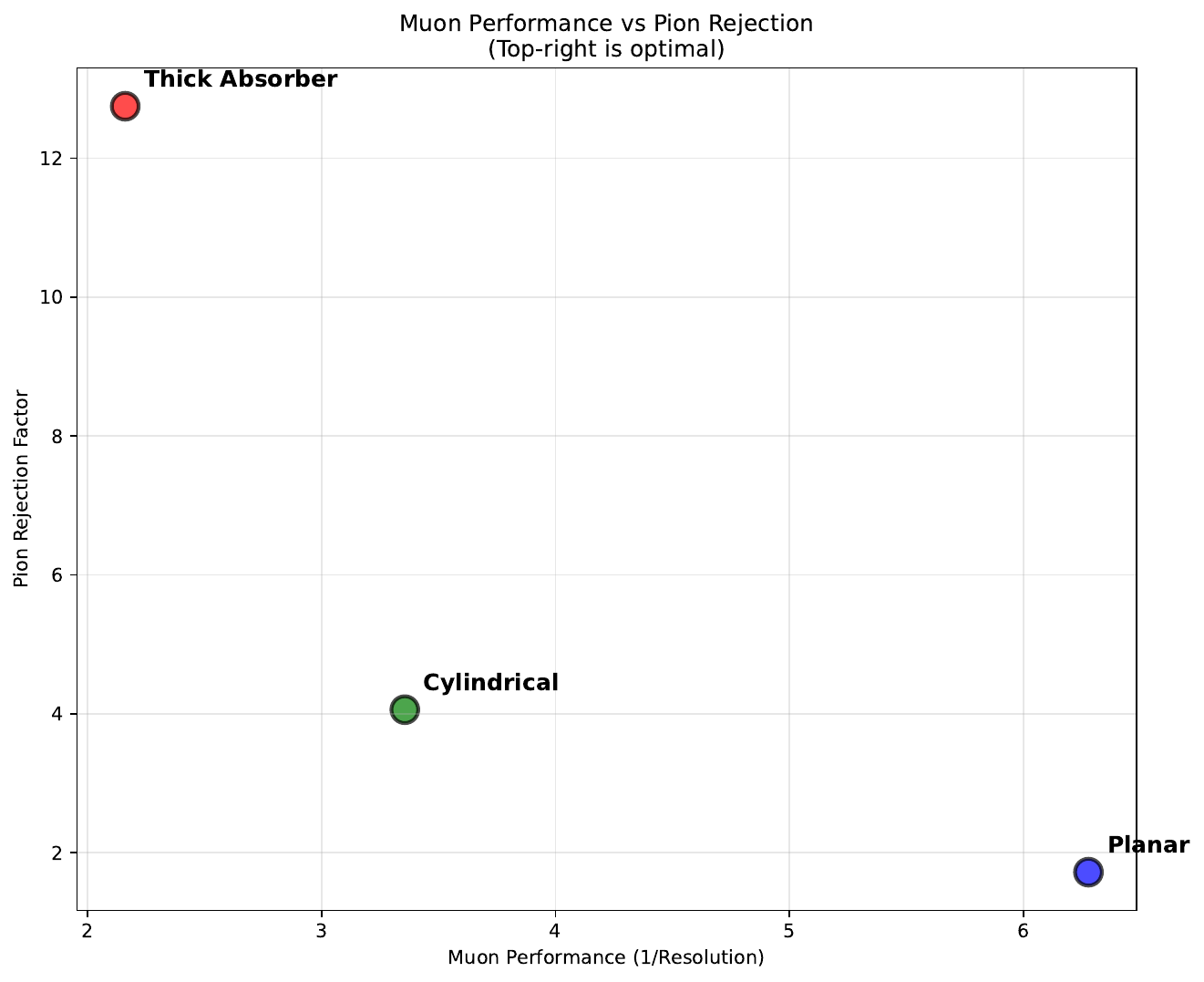}
    \caption{Muon performance versus pion rejection trade-off. The planar configuration (blue) optimizes muon precision, the thick absorber (red) optimizes pion rejection, and the cylindrical design (green) provides intermediate performance. No configuration achieves the optimal upper-right region, illustrating the fundamental physics trade-off.}
    \label{fig:muon_tradeoff}
\end{figure}

Figure~\ref{fig:muon_tradeoff} presents the fundamental trade-off between muon detection performance and pion rejection as a scatter plot, with muon performance (defined as 1/resolution, higher is better) on the x-axis and pion rejection factor on the y-axis. The optimal region lies in the upper right corner of this parameter space.

The three configurations occupy distinct regions: the planar design (blue) achieves the highest muon performance ($\sim$6.3) but lowest pion rejection ($\sim$1.7); the thick absorber (red) achieves the highest pion rejection ($\sim$12.75) but lowest muon performance ($\sim$2.2); and the cylindrical configuration (green) occupies an intermediate position ($\sim$3.4 muon performance, $\sim$4.1 pion rejection).

The clear anti-correlation between these metrics-no configuration achieves both high muon performance and high pion rejection-illustrates the fundamental physics trade-off that the agent correctly identified and quantified. Applications must select configurations based on their specific requirements: precision muon spectroscopy favors the planar design, while high-background environments requiring maximum hadron rejection favor the thick absorber approach.

\subsection{Summary of Key Metrics}

Table~\ref{tab:muon_summary} summarizes the quantitative performance metrics from the muon spectrometer design study.

\begin{table}[htbp]
    \centering
    \caption{Summary of muon spectrometer performance metrics across three configurations.}
    \label{tab:muon_summary}
    \begin{tabular}{@{}lccc@{}}
        \toprule
        \textbf{Metric} & \textbf{Planar} & \textbf{Cylindrical} & \textbf{Thick Absorber} \\
        \midrule
        Iron thickness per layer (cm) & 20 & 20 & 30 \\
        Total depth ($\lambda_I$) & 5.0 & 5.0 & 7.4 \\
        Muon detection efficiency & 100\% & 100\% & 100\% \\
        Muon resolution (average) & $0.159 \pm 0.008$ & $0.298 \pm 0.015$ & $0.463 \pm 0.023$ \\
        Muon resolution at 5 GeV & 0.0044 & 0.2653 & 0.1413 \\
        Muon resolution at 20 GeV & 0.1429 & 0.3146 & 0.4455 \\
        Muon resolution at 50 GeV & 0.3306 & 0.3141 & 0.5016 \\
        Pion rejection factor & $1.72 \pm 0.17$ & $4.06 \pm 0.41$ & $12.75 \pm 1.28$ \\
        Pion resolution (average) & 0.0434 & 1.9336 & 0.1763 \\
        \bottomrule
    \end{tabular}
\end{table}

\end{document}